\documentclass{emulateapj}
\usepackage{apjfonts}
\usepackage{natbib}

\newcommand{\chandra}{{\it Chandra\/}}
\newcommand{\flux}{{erg~cm$^{-2}$~s$^{-1}$}}
\newcommand{\lum}{{erg~s$^{-1}$}}

\begin{document}

\title{Identifications and Photometric Redshifts
of the 2 Ms {\bf {\em Chandra}} Deep Field-South Sources}
\author{
B.~Luo,\altaffilmark{1}
W.~N.~Brandt,\altaffilmark{1}
Y.~Q.~Xue,\altaffilmark{1}
M.~Brusa,\altaffilmark{2,3}
D.~M.~Alexander,\altaffilmark{4}
F.~E.~Bauer,\altaffilmark{5,6}
A.~Comastri,\altaffilmark{7}
A.~Koekemoer,\altaffilmark{8}
B.~D.~Lehmer,\altaffilmark{4,9,10}
V.~Mainieri,\altaffilmark{11}
D.~A.~Rafferty,\altaffilmark{1}
D.~P.~Schneider,\altaffilmark{1}
J.~D.~Silverman,\altaffilmark{12}
\& C.~Vignali\altaffilmark{13}
}
\altaffiltext{1}{Department of Astronomy \& Astrophysics, 525 Davey Lab,
The Pennsylvania State University, University Park, PA 16802, USA}
\altaffiltext{2}{Max-Planck-Institut f\"ur Extraterrestrische Physik,
Giessenbachstrasse, D-85748 Garching b. M\"unchen, Germany}
\altaffiltext{3}{University of Maryland, Baltimore County, 1000
Hilltop Circle, Baltimore, MD 21250, USA}
\altaffiltext{4}{Department of Physics, Durham University,
Durham, DH1 3LE, UK}
\altaffiltext{5}{Space Science Institute, 4750 Walnut Street, Suite
205, Boulder, Colorado 80301}
\altaffiltext{6}{Pontificia Universidad Cat\'{o}lica de Chile,
Departamento de Astronom\'{\i}a y Astrof\'{\i}sica, Casilla 306,
Santiago 22, Chile}
\altaffiltext{7}{INAF---Osservatorio Astronomico di Bologna, Via Ranzani 1,
Bologna, Italy}
\altaffiltext{8}{Space Telescope Science Institute, 3700 San Martin Drive,
Baltimore, MD 21218, USA}
\altaffiltext{9}{The Johns Hopkins University, Homewood Campus, Baltimore, MD
21218, USA}
\altaffiltext{10}{NASA Goddard Space Flight Centre, Code 662, Greenbelt, MD 20771, USA}
\altaffiltext{11}{European Southern Observatory, Karl-Schwarzschild-Strasse 2,
Garching, D-85748, Germany}
\altaffiltext{12}{Institute for the Physics and Mathematics of the Universe
(IPMU), University of Tokyo, Kashiwanoha 5-1-5, Kashiwa, Chiba 277-8568,
Japan}
\altaffiltext{13}{Universit\'a di Bologna, Via Ranzani 1, Bologna, Italy}

\begin{abstract}
We present reliable multiwavelength identifications and high-quality 
photometric redshifts for the 462 \hbox{X-ray} sources in 
the $\approx2$~Ms \chandra\ Deep \hbox{Field-South} survey. 
Source identifications are carried out using
deep \hbox{optical--to--radio} multiwavelength catalogs, and are then combined to
create lists of primary and secondary counterparts for the 
X-ray sources. We identified
reliable counterparts for 442 (95.7\%) of the X-ray sources, with an expected 
false-match probability of $\approx6.2\%$; we also
selected four additional likely counterparts.
The majority of the other 16 X-ray sources appear to be
off-nuclear sources, sources associated with galaxy groups and clusters,
high-redshift active galactic nuclei (AGNs), or spurious \hbox{X-ray} sources.
A \hbox{likelihood-ratio} method is used for source matching,
which effectively reduces the false-match probability at faint
magnitudes compared to a simple error-circle matching method.
We construct a master photometric catalog
for the identified X-ray
sources including up to 42 bands of \hbox{UV--to--infrared} data, and 
then calculate their photometric redshifts (\hbox{photo-z's}). 
High accuracy in the derived
photo-z's is accomplished owing to (1) the up-to-date
photometric data covering the full 
spectral energy distributions (SEDs) of the \hbox{X-ray} sources, (2)
more accurate photometric data as a result of source deblending for
$\approx10\%$ of the sources in the infrared bands and a few percent in the
optical and near-infrared bands, (3) a set of 265
galaxy, AGN, 
and galaxy/AGN hybrid templates carefully constructed
to best represent all possible SEDs, (4) the 
Zurich Extragalactic Bayesian Redshift Analyzer (ZEBRA) used to derive the 
photo-z's, which corrects the SED templates to
best represent the SEDs of real sources at different redshifts and thus
improves the photo-z quality. 
The reliability of the photo-z's is evaluated using the subsample
of 220 sources with secure spectroscopic redshifts. We achieve an
accuracy of $|\Delta z|/(1+z)\approx1\%$
and an outlier [with $|\Delta z|/(1+z)>0.15$] fraction of $\approx1$.4\%
for sources with spectroscopic redshifts. 
We performed blind tests to
derive a more realistic estimate of the photo-z quality for
sources without spectroscopic redshifts.
We expect there are $\approx9\%$ outliers for the 
relatively brighter sources ($R\la26$),
and the outlier fraction will increase
to $\approx15$--25\% for the fainter sources ($R\ga26$).
The typical \hbox{photo-z}
accuracy is \hbox{$\approx6$--7\%}.
The outlier fraction and photo-z
accuracy do not appear to have a redshift dependence (for $z\approx0$--4).
These photo-z's 
appear to be the best obtained so far for faint X-ray sources,
and they have been
significantly ($\ga50\%$) improved
compared to previous estimates of the photo-z's for the 
X-ray sources in the $\approx2$~Ms 
\chandra\ Deep Field-North and $\approx1$~Ms \chandra\ Deep Field-South.

\end{abstract}

\section{INTRODUCTION}

The \chandra\ Deep Field-North (CDF-N)
and \chandra\ Deep Field-South 
(\hbox{CDF-S}) are the two deepest \chandra\ surveys (see
\citealt{Brandt2005} for a review), each covering
$\approx440$ arcmin$^2$ areas with enormous multiwavelength 
observational investments. Together they
have detected about 1160 X-ray point sources (e.g., 
\citealt{Giacconi2002,Alexander2003,Luo2008}, hereafter L08),
with a sky density of $\approx10\,000$~deg$^{-2}$ at the limiting flux
of $\approx2\times10^{-17}$~\flux\ in the 0.5--2.0 keV band.
Most ($\approx75\%$) 
of these X-ray sources are active galactic nuclei (AGNs),
often obscured, at $z\approx0.1$--4; at faint fluxes, some of
the sources are starburst and normal galaxies at $z\approx0.1$--2
\citep[e.g.,][]{Bauer2004}.

The CDF-N and CDF-S provide an ideal opportunity to study 
AGN cosmological evolution, AGN physics, the role of AGNs 
in galaxy evolution, and the
X-ray properties of starburst and normal galaxies, groups and clusters
of galaxies, large-scale structures in the distant universe, and Galactic stars.
However, one prerequisite of such research is to associate the 
X-ray sources with their
correct counterparts at optical, near-infrared (NIR), infrared (IR),
and radio wavelengths, and then
to determine
their redshifts either spectroscopically or photometrically.
The CDF-S was recently expanded from $\approx1$~Ms to $\approx2$~Ms 
exposure, and 
more than 130 new X-ray sources are detected (L08). Meanwhile, 
additional deep 
multiwavelength surveys of the CDF-S have 
become available, including the 
VLA radio \citep[e.g.,][]{Kellermann2008,Miller2008}, {\it Spitzer} IR 
\citep[e.g.,][]{Damen2010}, NIR $K$-band
\citep[e.g.,][]{Grazian2006,Taylor2009}, and {\it GALEX} UV \citep[e.g.][]{Morrissey2005} surveys; 
there are also a few recent spectroscopic
surveys that provide updated redshift information 
\citep[e.g.,][]{Vanzella2008,Popesso2009}. 
It is thus crucial to provide
\hbox{up-to-date} X-ray source identifications in the CDF-S 
along with their redshifts to facilitate
\hbox{follow-up} scientific studies.

One major uncertainty about X-ray source identifications in deep surveys is
the contamination from spurious matches, e.g., 
because of the high source density
in deep optical catalogs. For example, \citet{Barger2003}
were able to assign optical counterparts to $\approx85\%$ 
of the X-ray sources in the $\approx2$~Ms CDF-N. However, they estimated 
the false-match probability to be $\approx10\%$ at $R\la24$, and 
it will rise to $\approx25\%$ at \hbox{$R\approx24$--26.} 
A simple error-circle method was used in their work, in which the closest 
optical source within a given matching radius of an X-ray source 
was selected as
the counterpart. Recently, a more sophisticated 
likelihood-ratio matching approach was adopted for matching X-ray sources in deep
\chandra\ and {\it XMM-Newton} 
surveys to optical/NIR sources \citep[e.g.,][]{Brusa2005,Cardamone2008,Laird2009,Aird2010}. 
The likelihood-ratio technique 
takes into account the positional
errors of both the optical and X-ray sources 
and also the expected magnitude distribution of the counterparts, and 
thus it can significantly reduce the false-match probability at faint 
magnitudes. Furthermore, performing the source matching at some
wavelengths other than optical, e.g., NIR or radio, can also reduce
the false-match probability, as the X-ray sources may be better associated 
with NIR or radio sources. Combining the multiwavelength X-ray source
identifications
will effectively lower the overall false-match probability.

As many X-ray sources in deep surveys are optically faint 
(and are therefore challenging targets to obtain spectroscopic redshifts),
a significant portion of 
the sources have their redshifts determined photometrically.
Photometric redshifts are 
derived by fitting spectral energy distribution (SED) templates
to the observed broadband photometric data, and their accuracy is largely
controlled by the quality of the data available. 
The previous photometric redshifts for the X-ray sources in the 
$\approx2$~Ms
CDF-N \citep[e.g.,][]{Barger2003,Capak2004} 
and $\approx1$~Ms CDF-S \citep[e.g.,][]{Wolf2004,Zheng2004}
are generally accurate to $\approx10\%$ with $\approx15$--25\% 
catastrophic redshift failures.
The superb and improved multiwavelength coverage in the CDF-S has
produced full SED sampling for the X-ray sources, from UV to IR,
allowing determination of photometric redshifts to significantly higher
accuracy.

In this paper, we present multiwavelength identifications of
the CDF-S X-ray sources in the main source catalog of L08.
Given the counterpart information, we compose a photometric catalog 
for these X-ray sources including up to 42 bands of UV--to--IR data.
Photometric redshifts are then calculated and compared to the 
latest secure spectroscopic redshifts to evaluate their accuracy. 
The identifications and spectroscopic redshifts
presented here supersede the counterpart and redshift information
provided in L08.
In \S2 we describe in
detail the likelihood-ratio matching method and our multiwavelength 
identification results. In \S3 we present the photometric catalog
and the derivation of photometric redshifts. The accuracy of the 
photometric redshifts is appropriately estimated. In \S4 we discuss
the nature of the unidentified X-ray sources, and present some
future prospects toward source identifications. We summarize in \S5.

Throughout this paper, all magnitudes are based upon the AB 
magnitude system \citep[e.g.,][]{Oke1983}, and we adopt the latest 
cosmology with
$H_0=70.5$~km~s$^{-1}$~Mpc$^{-1}$, $\Omega_{\rm M}=0.274$,
and $\Omega_{\Lambda}=0.726$
derived from the five-year WMAP observations \citep{Komatsu2009}.

\section{MULTIWAVELENGTH IDENTIFICATIONS OF THE 2 MS CDF-S X-RAY SOURCES}

\subsection{X-ray and Optical--to--Radio Data}

The X-ray sample we chose is  
the main \chandra\ source catalog for the $\approx2$~Ms 
\hbox{CDF-S}, which contains 462 X-ray sources (L08). These sources were
detected in three X-ray bands, 0.5--8.0 keV (full band), 0.5--2.0 keV (soft
band), and 2--8 keV (hard band), with a {\sc wavdetect} false-positive 
probability threshold of $1\times10^{-6}$. The resulting source lists
were then merged to create the main catalog (see \S3.3 of L08 for details). 
The survey covers an area of $\approx436$ arcmin$^{2}$ and reaches 
on-axis sensitivity limits of $\approx7.1\times10^{-17}$,
$\approx1.9\times10^{-17}$ and $\approx1.3\times10^{-16}$~\flux\
for the full, soft, and hard bands, respectively.

We have visually checked the positions of all the X-ray sources in L08
during an X-ray spectral extraction process (Bauer~et~al. 2010, in
preparation), using the $\approx2$~Ms CDF-S \chandra\ events file and sometimes
also the $\approx250$~ks Extended \chandra\ Deep Field-South (\hbox{E-CDF-S};
\citealt{Lehmer2005}) events file for sources with large off-axis angles.
Although most X-ray sources in L08 have good positions (no positional offsets
or offsets $<0.24\arcsec$), there are five sources having positions off 
by $>1\arcsec$.
These are sources 62, 90, 125, 338, and 456, and we shifted their positions by
1.2\arcsec, 3.2\arcsec, 1.3\arcsec, 2.8\arcsec, and 1.5\arcsec, respectively.
Sources 90 and 456 are located at large off-axis angles (11.8\arcmin\ and 
9.8\arcmin); the other
three sources are faint ($\approx21$--26 detected counts
in the soft band
with errors of about 8--10 counts), and might be
composite sources (blended with a weak nearby object).
The complex point spread functions or morphologies of these five sources
make it difficult to
determine the centroid positions, and their positions reported in L08
are poorly defined. \footnote{Four of these sources (all but source
456) do not have any 
apparent counterparts at their original positions.}

We searched for counterparts of the
$\approx2$~Ms \hbox{CDF-S} \hbox{X-ray} sources in several deep 
\hbox{optical--to--radio} 
catalogs, most of which cover the entire \hbox{CDF-S} region.
Due to dust absorption and redshift effects, \hbox{X-ray} sources 
in deep surveys usually
appear brighter at redder wavelengths.\footnote{Counterpart searching has
also been performed at several shorter wavelengths (e.g., $V$ and $U$ bands),
which yield fewer identifications.}
Therefore we used the following
catalogs 
for identification purposes.
\begin{enumerate}

\item
The WFI $R$-band catalog \citep{Giacconi2002,Giavalisco2004},
with a 5 $\sigma$ limiting AB magnitude of
$27.3$.

\item
The GOODS-S {\it HST} version r2.0z $z$-band catalog \citep{Giavalisco2004}\footnote{See 
\url{http://archive.stsci.edu/pub/hlsp/goods/catalog\_r2/}\,. 
The r2.0z catalog is based on the v2.0 GOODS-S images, 
which have significantly longer total exposure times in the $z$ band
than the v1.0 images.}, with
a 5 $\sigma$ limiting AB magnitude of
$28.2$.
The GOODS-S survey covers a solid angle of $\approx160$~arcmin$^2$ 
in the center of the \hbox{CDF-S}.

\item
The GEMS {\it HST} $z$-band catalog \citep{Caldwell2008},
with a 5 $\sigma$ limiting AB magnitude of
$27.3$. The GEMS survey was designed to exclude the GOODS-S region, and thus
does not cover the entire CDF-S. 
However, \citet{Caldwell2008} reduced the GOODS-S v1.0 data 
identically to the reduction of GEMS, and the published $z$-band catalog
covers the entire CDF-S.\footnote{We used the entire GEMS catalog in the 
source-identification process for consistency. However, GEMS counterparts
have a lower priority than GOODS-S counterparts when we select primary
counterparts for the X-ray sources (see \S\ref{results})} 

\item
The GOODS-S MUSIC $K$-band catalog \citep{Grazian2006},
with a $90\%$ ($\approx1.8~\sigma$) limiting AB magnitude of
$23.8$. The GOODS-S MUSIC survey covers a solid angle of $\approx140$~arcmin$^2$ 
in the center of the CDF-S. The V2.0 catalog was used 
(A. Grazian 2009, private communication).

\item
The MUSYC $K$-band catalog \citep{Taylor2009},
with a 5 $\sigma$ limiting AB magnitude of $22.4$. This catalog is not as 
deep as the GOODS-S MUSIC $K$-band catalog, but it covers the entire CDF-S.

\item
The SIMPLE 3.6 $\mu$m catalog \citep{Damen2010}, with a
5 $\sigma$ limiting AB magnitude of
23.8.\footnote{We also searched for SIMPLE 4.5~$\mu$m counterparts, which
yielded fewer identifications than the 3.6 $\mu$m catalog.}

\item
The VLA 1.4 GHz radio catalog from \citet{Miller2008}. We used the 5~$\sigma$ 
catalog provided by N.A. Miller (2009, private communication),
with a
5 $\sigma$ limiting flux density of $\approx 40~\mu$Jy.
Compared to the 
optical positions (WFI $R$ band or GOODS-S $z$ band), the radio positions have 
small systematic offsets in right ascension and declination; since 
the X-ray positions were registered to the optical positions (L08), we manually
shifted all the radio positions by $-0.18$\arcsec\ in right ascension and
0.28\arcsec\ in declination to remove the systematic offsets.

\end{enumerate}

\begin{figure}
\centerline{
\includegraphics[scale=0.5]{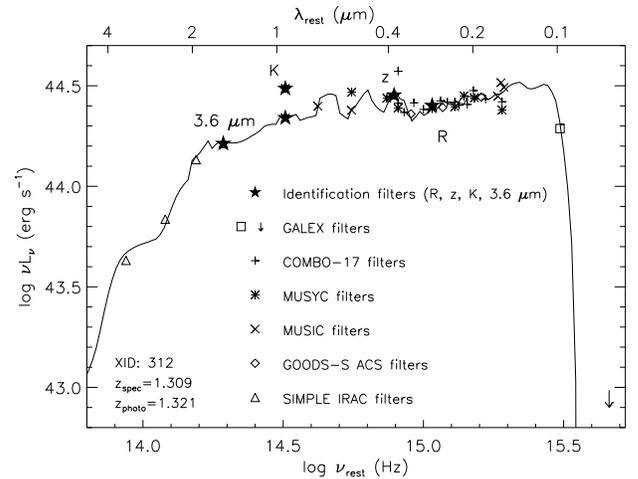}
}
\figcaption{
The SED of source 312 in L08 along with the
selected best-fit SED template (a composite starburst/spiral galaxy template), 
as an example to illustrate the photometric bands
used for source identification, the coverage of the broadband 
photometric data, and the photometric redshift fitting result.
Error bars for the data points are smaller than or comparable to
the symbol sizes and thus are not shown.
Although this source is not detected in the radio band, the VLA 1.4 GHz
radio catalog 
was also used in the identification process (not shown in this plot). 
The source is not detected in
the {\it GALEX} \hbox{far-UV} band, and the $1~\sigma$ upper limit is shown. 
There are 42 filters of photometric data displayed in the SED.
Note that not all of the data points are used 
in the photometric redshift calculation (see \S\ref{phodata} for details).
The discrepancy between some photometric data from different catalogs
is probably caused by AGN variability.
\label{sedexp}}
\end{figure}

The basic optical/NIR/IR/radio
(hereafter ONIR) catalog information is listed in Table~\ref{matchsum}. 
Note that the GOODS-S and MUSIC catalogs do not cover the entire CDF-S.
We also
include in Table~\ref{matchsum} 
the number of sources in each of these catalogs that are within the $\approx2$ 
Ms CDF-S region ($N_{\rm o}$).
The locations of the optical, NIR, and IR bands in a sample SED are shown in
Figure~\ref{sedexp}.

\subsection{Matching Method}\label{method}
Source matching to the X-ray catalog was performed for each of the 
optical--to--radio
catalogs described above.
We used the likelihood-ratio technique to identify the
ONIR
counterparts of the \hbox{X-ray} sources \citep[e.g.,][]{Sutherland1992, 
Ciliegi2003,Brusa2005,Brusa2007}. 
Due to the high source densities of the ONIR
catalogs and centroid errors in the X-ray positions, 
the simple method of searching for the
nearest counterpart within a small radius could yield a non-negligible number
of false matches, usually to faint ONIR sources where
the background source density is high.
The likelihood-ratio technique (described in detail below) 
takes into account the positional 
accuracy of both the 
ONIR and \chandra\ catalogs and also the expected magnitude distribution 
of the counterparts. It assigns a reliability parameter to all the 
possible counterparts, and it mitigates the effect of false matches to faint 
ONIR sources.

For an ONIR candidate with a magnitude $m$ (at a given band)
at
an angular separation $r$ from a given \hbox{X-ray} 
source,\footnote{The AB magnitudes
for radio sources were converted from the radio flux densities 
\citep[e.g.,][]{Oke1983}, 
\hbox{$m({\rm AB})=-2.5\log(f_\nu)-48.60$}, where
$f_\nu$ is the flux density in units of erg~cm$^{-2}$~s$^{-1}$~Hz$^{-1}$.} 
the likelihood ratio is defined as the ratio of the
probability of the ONIR object being the true counterpart of the X-ray source
and the corresponding 
probability of the ONIR object being 
a background, unrelated object \citep[e.g.,][]{Ciliegi2003},
\begin{equation}
LR=\frac{q(m)f(r)}{n(m)}~,
\end{equation}
where $q(m)$ is the expected magnitude distribution of counterparts, $f(r)$
is the probability distribution function of the angular separation, and $n(m)$
is the surface density of background objects with magnitude $m$. 
We assume
that the probability distribution of angular separations follows a
Gaussian distribution \citep[e.g.,][]{Ciliegi2003}:
\begin{equation}
f(r)=\frac{1}{2\pi\sigma^2}\exp{\left(\frac{-r^2}{2\sigma^2}\right)}~,
\end{equation}
with the standard deviation $\sigma=\sqrt{\sigma_{\rm o}^2+\sigma_{\rm X}^2}$, 
where $\sigma_{\rm o}$ 
and $\sigma_{\rm X}$ are the 1 $\sigma$ positional errors of the ONIR
and \hbox{X-ray} sources, respectively. 
The 1 $\sigma$ positional errors
of the \hbox{X-ray} sources were estimated by correlating \hbox{X-ray} sources
to bright WFI \hbox{$R$-band} sources and then examining the positional offsets
(see \S3.3.1 of L08 for details). 
The 1 $\sigma$ positional errors of the
ONIR sources are either
provided by the catalog or estimated
from the distributions of positional
offsets
by matching the
sources to the GOODS-S $z$-band catalog. The ONIR positional errors
and X-ray positional errors are listed in Tables~\ref{matchsum} and 
\ref{cat}, respectively.

\begin{figure}
\centerline{
\includegraphics[scale=0.5]{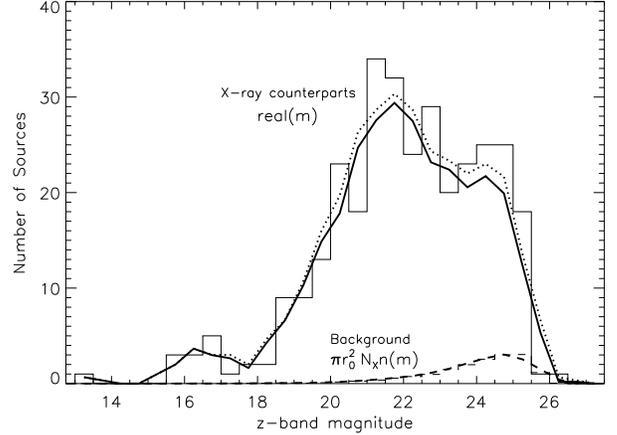}
}
\figcaption{The expected magnitude distributions of the background sources
(thin dashed histogram; normalized by a factor of $\pi r_0^2N_{\rm X}$) 
and counterparts (thin solid histogram) for 
the GEMS $z$-band catalog in the first trial run. The 
counterparts in the first trail run are GEMS sources
within $1\arcsec$ of any \hbox{X-ray} source, and they were only 
used to calculate $q(m)$ and thus $LR$ (see \S\ref{method}).
The thick solid and dashed curves are the corresponding smoothed 
(with a boxcar average)
distributions that represent ${\rm real}(m)$
and $\pi r_0^2N_{\rm X}n(m)$ (see \S\ref{method} for details). The thick
dotted curve shows the final smoothed magnitude distribution of the 
counterparts after three iterations. The magnitude distribution of 
the background sources peaks at fainter magnitude than the distribution
of counterparts.
\label{gemszhist}}
\end{figure}

The magnitude-dependent surface density of the background sources, $n(m)$, 
was estimated
using a sample of ONIR sources that are at a angular distance between
$5\arcsec$ and $30\arcsec$
from any of the \hbox{X-ray} sources. We chose sources inside such annular regions to 
exclude possible counterparts from the 
background calculation. 
The expected magnitude distribution of counterparts $q(m)$ is not directly
observable. To estimate $q(m)$, 
we ran the matching procedure in an iterative way. The matching results from a
previous run were used to calculate $q(m)$ for the next run.
As a first guess, we  
selected all ONIR sources 
within $1\arcsec$ of any \hbox{X-ray} source, and 
we assume these ONIR objects 
are representative of the counterparts for the X-ray sources 
with a small contribution
from coincident background sources.
The number of sources in this sample is $N_1$, and the 
magnitude distribution of these sources is denoted as ${\rm total}(m)$, which
was then background subtracted to derive
\begin{equation}
{\rm real}(m)={\rm total}(m)-\pi r_0^2N_{\rm X}n(m)~,
\end{equation}
where $N_{\rm X}$ is the total number of \hbox{X-ray} sources in the relevant 
sample (see Col.~9 of Table~\ref{matchsum}), and 
$r_0=1\arcsec$. The background subtraction generally has 
a small effect (a few percent) on the resulting magnitude distribution.
See Figure~\ref{gemszhist} for an example of 
${\rm real}(m)$ and $n(m)$ in the GEMS $z$-band catalog.
Due to the magnitude limits of the ONIR catalogs,
we are only able to detect a fraction $Q$ of all the true counterparts 
($Q=\int_{-\infty}^{m_{\rm lim}}q(m)~dm$). Thus the expected magnitude distribution 
of counterparts $q(m)$ is derived by normalizing ${\rm real}(m)$ 
and then multiplying by $Q$:
\begin{equation}
q(m)=\frac{{\rm real}(m)}{\sum_{m}{\rm real}(m)}Q~.
\end{equation}
The factor $Q$ was first set to 
$Q=N_1/N_{\rm X}$. The magnitude distributions, $q(m)$ and $n(m)$, 
are smoothed using a boxcar
average (with width of one magnitude) 
and then used as input in the likelihood-ratio calculation.
We note that source blending, which is most severe in the 3.6~$\mu$m
catalog, could affect the values of $q(m)$ and $n(m)$ by a tiny fraction and
thus introduce small uncertainties in the matching results. However, as we choose
counterparts carefully combining all seven catalogs (see \S2.3),
this will not affect our final identifications.

Having computed the values of $q(m)$, $f(r)$, and $n(m)$, we calculated $LR$
for all the ONIR sources within $5\arcsec$ of any of the 
X-ray sources. We then
chose a threshold value for $LR$ ($L_{\rm th}$) to discriminate between
spurious and real counterparts. An ONIR source is considered to be 
a counterpart for an \hbox{X-ray} source if its $LR$ value exceeds $L_{\rm th}$.
In some cases, there will be more than one counterpart for an 
\hbox{X-ray} source within our $5\arcsec$ limiting radius. 
If the ratio between the highest
and the second highest $LR$ values is greater than
5, we keep only the counterpart with the highest $LR$ value\footnote{We
believe those counterparts with much smaller $LR$ values are coincident
matches; about 16 such unlikely counterparts were identified and removed 
in total, considering
all the optical--to--IR catalogs.
Because the final counterpart 
for an X-ray source is selected after combining the matching results
for all the catalogs, 
the chance of removing a possible true counterpart
here is negligible.};
otherwise, the \hbox{X-ray} source is considered to have multiple 
counterparts in this catalog.
For each counterpart
for an \hbox{X-ray} source identified this way,
we calculate a reliability parameter, $R_{\rm c}$, which represents
the probability of it being the correct identification:
\begin{equation}
R_{\rm c}=\frac{LR}{\sum LR+(1-Q)}~,
\end{equation}
where the sum is over all the counterparts for this
\hbox{X-ray} source.

The choice of $L_{\rm th}$ is not unique, and it depends on two 
factors: first, it should be small enough to avoid missing many real 
identifications; second, it should be large enough to prevent 
having many spurious identifications. 
In this paper, we followed the discussions in 
\citet{Ciliegi2003}, \citet{Brusa2005,Brusa2007}, and 
Civano~et al. (2010, in preparation),
and we chose a $L_{\rm th}$ that
maximizes the sum of sample completeness and reliability.
The sample 
completeness parameter ($C$) is defined as the ratio between the sum of the 
reliability of all the sources identified as counterparts and the total number
of \hbox{X-ray} sources, $C=\sum R_{\rm c}/N_{\rm X}$. 
The sample reliability parameter ($R$) is defined as the 
average of the reliabilities of all the sources identified as 
counterparts. For each run of the likelihood-ratio method, we tried several
values of $L_{\rm th}$ and chose the one which maximizes the sum of 
$C$ and $R$. 

After
the first trial run of the likelihood-ratio procedure, we calculated ${\rm real}(m)$ 
as the magnitude distribution of all the sources identified as counterparts
in the previous run, and we approximated the normalization factor $Q$ to be
the completeness $C$ of the previous run. For each ONIR 
catalog, the likelihood-ratio procedure was run iteratively until
the resulting values of $C$ and $R$ converged. This is a stable
process with mild perturbations, and the results converge quickly, usually in
$\approx2$--5 iterations.

\subsection{Matching Results}\label{results}

A summary of the likelihood-ratio matching parameters and results is shown in 
Table~\ref{matchsum}. We list the total number of \hbox{X-ray} sources
that are within the coverage of the ONIR catalog ($N_{\rm X}$),
the number of \hbox{X-ray} sources identified with at least one ONIR
counterpart ($N_{\rm ID}$), the number of \hbox{X-ray} sources without
identifications ($N_{\rm NoID}$), and the number of \hbox{X-ray} sources 
identified with at least two ONIR counterparts ($N_{\rm Multi}$).
We estimated the false-match probability of these identifications 
using a bootstrapping approach.
We shifted the \hbox{X-ray} source coordinates by 5\arcsec--53\arcsec\
(with a 2\arcsec\ increment between shifts) in right ascension and 
declination respectively, 
and recorrelated the shifted sources 
with the ONIR sources using the likelihood-ratio
method. This was done for 100 trials,
and the results were averaged to estimate the number of false matches. 
The expected number of false matches ($N_{\rm False}$) and the 
false-match probability for each catalog are shown in Table~\ref{matchsum}.
The false-match probabilities range from $\approx1$--9\%.
The matching to the radio sources yields very secure counterparts, 
with less than one expected false match, largely due to the small
background radio source density.

For each catalog, we calculated its counterpart recovery rate, 
defined as the expected number of
true counterparts ($N_{\rm ID}-N_{\rm False}$) divided by the number of
X-ray sources ($N_{\rm X}$), shown in Table~\ref{matchsum}. 
Compared to the completeness parameter $C$, the recovery rate parameter
utilizes the bootstrapping results, and is a more appropriate estimate
of the catalog completeness.
The
MUSIC NIR and SIMPLE IR catalogs have the highest recovery rates, while
the VLA radio catalog recovers the fewest counterparts.
We also calculated an \hbox{X-ray--to--ONIR} fraction (X--O fraction) parameter, 
defined as the expected number of 
true counterparts ($N_{\rm ID}-N_{\rm False}$) divided by the number of
ONIR sources ($N_{\rm o}$) in the catalog. The X--O fraction indicates
the degree of association between the X-ray and ONIR sources. 
Only $\approx1$--4\%
of the optical/NIR/IR sources are X-ray sources, whereas $\approx28\%$
of the radio sources are detected by \chandra.
A larger X--O fraction will lead to more reliable source matching results 
because of the lower background source density.
Both the recovery rate
and X--O fraction parameters are dependent on the catalog depth.

\begin{figure}
\centerline{
\includegraphics[scale=0.5]{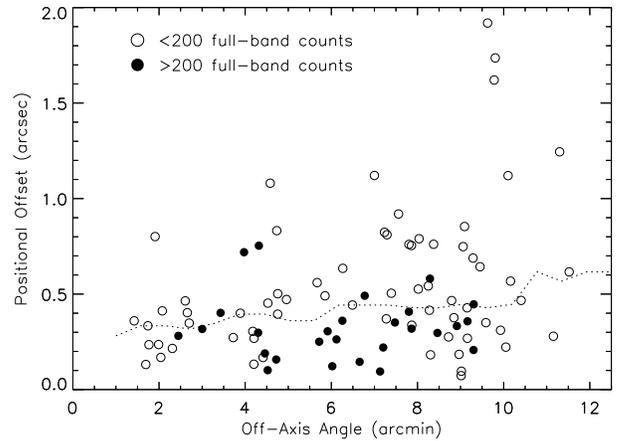}
}
\figcaption{
Positional offset vs. off-axis angle for
sources in the main X-ray source catalog of L08 that were matched to the VLA
radio sources \citep{Miller2008}.
The open circles represent \chandra\ sources with $<200$
\hbox{full-band}
counts, and the solid dots are \chandra\ sources with $\ge200$
\hbox{full-band} counts. The dotted curve shows the running median of all sources in
bins of 2\arcmin. The three outliers on the top right corner of the plot 
are faint X-ray sources with large positional errors 
($\sigma_{\rm X}\approx0.5\arcsec$--0.7\arcsec).
\label{dposradio}}
\end{figure}

Given the excellent radio-source positional accuracy and the secure matching
results, 
we can investigate the accuracy of the X-ray source positions using the 
information for the radio 
counterparts.\footnote{Three of the radio counterparts (those of sources
200, 364, and 368) are identified as 
the lobe components of multi-component radio sources by \citet{Kellermann2008}.
However, the radio counterparts of sources 364 and 368 are probably two
interacting galaxies based on optical imaging and spectroscopic results 
\citep[e.g.,][]{Mainieri2008}; these two radio sources are 
blended in \citet{Kellermann2008}, but well resolved in \citet{Miller2008}.}
In Figure~\ref{dposradio}, we show
the positional offset between the 94 X-ray sources and their radio
counterparts versus off-axis angle. The median offset is $0.4\arcsec$,
and there are clear off-axis angle and source-count dependencies.
The off-axis angle dependence is mainly due to the degradation of
the \chandra\ PSF
at large off-axis angles, while the count dependence is due to the
difficulty of finding the centroid of a faint \hbox{X-ray} source.
Although the results are comparable to the L08 study of the positional accuracy
based on the matching to the WFI $R$-band sources (see \S3.3.1 of L08),
Figure~\ref{dposradio} should be considered more 
representative of the typical X-ray
positional accuracy, as the radio catalog provides the most-reliable 
X-ray source identification.

Twenty of the 462 \hbox{X-ray} sources do not have a counterpart
in any of the ONIR databases,
based on the likelihood-ratio matching to all seven 
catalogs.
Four (XIDs 72, 165, 234, and 396) of these 20 sources have a faint ONIR source
(a $z$-band source in each case, with $z=25.3$--26.7)
within $\approx0.7$--1.2\arcsec\ of the \hbox{X-ray} position;
these ONIR sources were not selected by the \hbox{likelihood-ratio} matching due to
their small likelihood ratios ($LR<L_{\rm th}$).
However, given the relatively small positional offsets of these four sources,
the lack of any other apparent counterparts, 
and the fact that the likelihood-ratio matching is a statistical method and 
cannot possibly identify all the counterparts,
we manually assign them as the counterparts of the relevant
X-ray sources (see Col.~13 of Table~\ref{cat}). 
Thus 446 \hbox{X-ray} sources have identifications, including these four 
additional likely cases. 
Note that the \hbox{false-match} probabilities in Table~1 do not apply to these 
four sources.
For the remaining 16 sources without identifications, there is 
no ONIR source within $1.3\arcsec$. 
As the ONIR catalogs we used were all generated using red bands,
we also visually inspected images in bluer bands (e.g., $U$, $B$, and $V$)
at the positions of these 
16 sources, 
and verified that we did not miss any apparent blue counterparts for them.
Eight of these unidentified sources are probably related to some other objects 
(e.g., off-nuclear sources; see Col. 13 of Table \ref{cat}), and the majority
of the other eight are likely to be spurious X-ray sources (see discussion
in \S\ref{noiddis}).

Twenty of the sources only have a SIMPLE counterpart; however, eight of these
SIMPLE counterparts have a corresponding optical/NIR source nearby 
(within 0.8\arcsec), and they are likely to be the same source. 
Therefore we used the more accurate optical/NIR positions for these
eight counterparts (see Col.~13 of Table~\ref{cat}).

Some X-ray sources could have multiple counterparts, either from one 
ONIR catalog as a result of the likelihood-ratio matching (e.g., see
the $N_{\rm Multi}$ parameter in Table~\ref{matchsum}), or from 
different optical--to--radio catalogs.
There are 12 \hbox{X-ray} sources with multiple counterparts in at least
one of the ONIR catalogs; none of them has more than two
counterparts in one band. For the remaining 434 sources, we
consider an \hbox{X-ray} source to
have multiple counterparts if the angular 
distance between two of the counterparts in 
different bands exceeds $3\sqrt{\sigma_1^2+\sigma_2^2}$ (so that they appear
to be different sources), where $\sigma_1$
and $\sigma_2$ are the positional errors in the relevant ONIR catalogs.
There are 60 such sources; five of them have three counterparts. 
In all, 72 \hbox{X-ray}
sources have multiple counterparts (with the other 374 X-ray sources having 
a unique counterpart), 
and they are flagged in Column~12 of Table~\ref{cat}.

For each of the 446 X-ray sources with an identification,
we visually inspected all its counterparts in the appropriate bands, 
and chose a primary one. 
The position of this source
is the most probable position of the counterpart. 
We chose the primary counterpart from, in
order of priority, the VLA, GOODS-S, GEMS, MUSIC, WFI, MUSYC, or
SIMPLE catalog. This order is chosen based on several related factors: 
the positional accuracy,  
angular resolution (to minimize any blending effects),
false-match probability, and catalog depth.\footnote{
Although we adopted the same positional error for some catalogs (e.g., GOODS-S
and GEMS), the deeper catalogs (e.g., GOODS-S) can in principle provide 
slightly better positions for the counterparts.}
There are a few exceptions: (1) if the optical counterpart
is clearly extended, i.e., a nearby galaxy, 
we select the optical position instead of the VLA radio
position because the radio emission could arise from an off-nuclear structure 
(7 cases); 
(2) if the optical counterpart is bright and significant, while 
the radio counterpart is detected at a smaller than 7~$\sigma$ significance,
we prefer the optical position (4 cases); (3) when the counterpart from
a lower-priority catalog is much brighter and better resolved, or it has a 
significantly larger reliability parameter, we prefer this source as the 
primary counterpart (5 cases).
The final number of primary 
counterparts selected in each catalog ($N_{\rm Pri}$) is listed 
in Table~\ref{matchsum}.
The mean \hbox{false-match} probability for the 442 primary counterparts 
(excluding the the four
additional likely counterparts) is then 
estimated by taking the average of the false-match probabilities of 
individual catalogs weighted by the number of primary counterparts in each 
catalog. The mean false-match probability is $\approx6.2\%$, and it 
represents the identification reliability for the entire X-ray sample 
(excluding the the four additional likely counterparts).
The mean false-match probability also has slight off-axis angle and source-count
dependencies (due to its relation to the X-ray positional accuracy); 
e.g., it is $\approx6.0\%$ for sources within 4\arcmin,
$\approx7.0\%$ at $>8\arcmin$, $\approx9.8\%$ for sources with $<30$ full-band
counts, and $\approx5.8\%$ for $>30$ full-band counts.

A list of the primary counterparts for the 446 \hbox{X-ray} sources 
(out of 462 total sources) is presented in Table~\ref{cat}, 
with the details of the columns given below; 
we also list all the secondary counterparts (for sources with multiple counterparts) 
in Table~\ref{cat2nd} with a 
similar format.

\begin{itemize} 

\item
Column~(1): the \hbox{X-ray} source ID number, which is the same as the source
number in the main catalog of L08.

\item
Column~(2): The 1 $\sigma$ \hbox{X-ray} source positional error $\sigma_{\rm X}$. 

\item
Column~(3): the ONIR ID number of the primary counterpart, 
taken from the corresponding ONIR catalog (see Col.~10).
For the 16 \hbox{X-ray} sources that lack a counterpart, the
value is set to ``0''.

\item Columns~(4) and (5): the ONIR J2000 right ascension and declination of
the counterpart, respectively. This position is taken from the corresponding 
ONIR catalog (see Col.~10). \hbox{X-ray} sources that do not 
have a counterpart have this column set to ``0''.

\item Columns~(6): the approximate 1~$\sigma$ positional error of the 
counterpart, in units of arcseconds, 
taken from Column~6 of Table~\ref{matchsum}.

\item
Column~(7): the angular distance between the \hbox{X-ray} source and its ONIR
counterpart, in units of arcseconds, set to ``0'' if there is no counterpart.

\item
Column~(8): the angular distance divided by the quadratic sum of the positional errors (\hbox{$\sigma=\sqrt{\sigma_{\rm o}^2+\sigma_{\rm X}^2}$}), i.e., 
the angular distance in units of $\sigma$.

\item
Column~(9): the reliability parameter of this counterpart (see Eq. 5).
\hbox{X-ray} sources that do not have a counterpart have a value
of ``0''.

\item
Column~(10): the ONIR catalog from which the primary counterpart was
selected. 
\hbox{X-ray} sources that do not have a counterpart have this column 
set to ``$\cdots$''.

\item
Column~(11): the ONIR magnitude of the primary counterpart (see
Col.~10 for the band in which the magnitude was measured).
\hbox{X-ray} sources that do not have a counterpart have this column
set to ``0''.

\item
Column~(12): a flag on whether the \hbox{X-ray} source has multiple 
ONIR counterparts, set to ``1'' for true and ``0'' for false.

\item
Column~(13): notes on some special sources, including the 
16 \hbox{X-ray} sources that do not have 
ONIR identifications (cases {\it a}--{\it e})
and a few other sources (cases {\it f}--{\it h}). 
({\it a}) NoID: there is no ONIR source within $\approx1.3$--3.6\arcsec\ of the 
\hbox{X-ray} source position, and it is likely that this \hbox{X-ray} source
does not have an ONIR counterpart. There are eight such sources.
({\it b}) Edge: this source lies at the edge of the $\approx2$~Ms \hbox{CDF-S} field,
with a large PSF size. The X-ray photons could come from any of a number of
galaxies in the vicinity. There is one such source. 
({\it c}) Cluster: the \hbox{X-ray} source may 
be associated with the diffuse emission from a galaxy group or cluster. 
Galaxy groups and clusters in the $\approx2$~Ms CDF-S are
identified through a wavelet detection of extended X-ray emission
(Finoguenov et al. 2010, in preparation).
There are four such sources, which are generally faint and weakly detected
in the X-ray.
({\it d}) Star: the source is only detected in the soft X-ray 
band, 
and is 2.7\arcsec\ away from a bright ($R=16$) star.
The X-ray photons are probably largely from the star, 
and the X-ray position was not correctly determined by {\sc wavdetect}
due to the contamination of another possible faint X-ray source 
nearby.
There is one such source. 
({\it e}) ULX: this is an off-nuclear X-ray source associated
with a nearby galaxy. There are two such sources. One of them (source 301)
is reported in \citet{Lehmer2006}. 
({\it f}) Blend: the primary counterpart is 
blended with an adjacent source, and the position of the counterpart 
is poorly determined (positional uncertainty $\approx1$--2\arcsec).
There is one such source, and the primary counterpart is
from the SIMPLE catalog (and there are no detections of this source at 
shorter wavelengths). 
({\it g}) Faint: there is at least one faint ONIR
source within $\approx0.7$--1.2\arcsec\ of the \hbox{X-ray} 
source position. The likelihood
ratio parameter of this ONIR source is below the threshold,
but we manually assigned it to be the counterpart. There are four such sources.
({\it h}) Position: the source only has a counterpart 
from the SIMPLE catalog. However, the SIMPLE counterpart likely coincides 
with a nearby optical/NIR source, and we choose this optical/NIR source
as the primary counterpart because of its more accurate position (see above). 
There are eight such sources.

\end{itemize}

\begin{figure}
\centerline{
\includegraphics[scale=0.4]{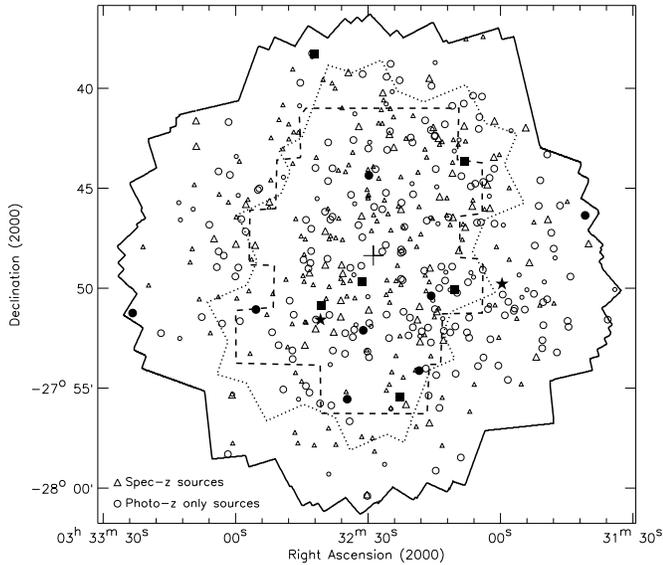}
}
\figcaption{Positions of the 446 identified (open symbols)
and 16 unidentified (filled symbols) X-ray sources. The open triangles
and circles represent sources with spectroscopic and photometric redshifts,
respectively, and small (large) symbols indicate redshifts smaller (larger)
than 1.3 (see \S\ref{zresults} for details about redshifts). The eight filled
circles represent sources that are probably related to a star,
galaxy groups and clusters, off-nuclear
sources, or any of a number of galaxies at the edge
of the CDF-S field (see Col. 13 of Table \ref{cat}). The six filled squares
and two filled stars represent the ``NoID'' (Col. 13 of Table \ref{cat}) 
sources that were detected at a {\sc wavdetect} false-positive
threshold of $1\times10^{-6}$ and $1\times10^{-8}$, respectively (see
\S\ref{noiddis} for discussion).
The solid, dotted, and dashed curves outline the regions covered by the 
$\approx2$~Ms \hbox{CDF-S}, GOODS-S, and MUSIC surveys, respectively.
The cross near the center indicates the average aim point of the CDF-S.
\label{coverage}}
\end{figure}

\begin{figure}
\centerline{
\includegraphics[scale=0.5]{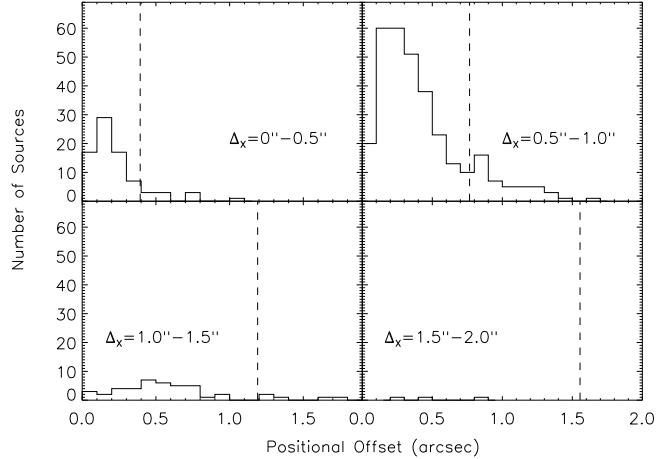}
}
\figcaption{
Histograms showing the distributions of the positional offset
between an X-ray source and its primary counterpart.
X-ray sources are
divided into four bins based on their $\approx85\%$ positional uncertainties ($\Delta_{X}$)
given in L08: $0\arcsec$--$0\farcs5$, $0\farcs5$--$1\farcs0$,
$1\farcs0$--$1\farcs5$, and $1\farcs5$--$2\farcs0$.
The vertical dashed lines indicate the median positional uncertainties (L08) for
X-ray sources in each bin.
The positional offsets are generally
smaller than or comparable to the X-ray $\approx85\%$
positional uncertainties for most
of the sources.
\label{dpos}}
\end{figure}

The locations of the 446 identified and 16 unidentified \hbox{X-ray} sources in
the CDF-S region are shown in Figure~\ref{coverage}, along with the coverages
of the GOODS-S and MUSIC catalogs, which are the only two catalogs that do not
cover the entire CDF-S.
For the 446 sources with an ONIR identification, the mean and median
positional offsets between the X-ray sources and their primary counterparts
are 0.41\arcsec\ and 0.32\arcsec, respectively. The maximum
offset is $\approx1.9\arcsec$.
Several X-ray
sources have large positional offsets ($\ga 1.5\arcsec$) to their primary counterparts.
These counterparts are either selected from the SIMPLE catalog
(less-accurate positions), VLA catalog (low background source density),
or are relatively bright, so that the likelihood ratio could still exceed
the threshold
at large ONIR--X-ray separations (see \S\ref{method}).
Figure~\ref{dpos} shows 
the distributions of the positional offsets in four bins of different 
X-ray positional uncertainties. The positional offsets are generally
smaller than or comparable to the $\approx85\%$ X-ray positional uncertainties
(dashed lines in Fig.~\ref{dpos}) for most sources. 

\subsection{Comparison with the Error-Circle Matching Method}

To check the efficiency of the likelihood-ratio technique, we compared
the matching results above to those derived from the simple 
error-circle method described in \S3.3.1 of L08.\footnote{In L08, 
we searched for the closest optical source
within a matching radius of $1.5\sqrt{\Delta_{\rm X}^2+\sigma_{\rm o}^2}$,
where $\Delta_{\rm X}$ is the $\approx85\%$ confidence-level
X-ray positional error.}
For the WFI $R$-band, L08 gives 347 matches
with an expected false-match probability of $12\%$ (305 expected
true matches), while here we have
344 matches
with a false-match probability of $9\%$ (312 expected
true matches).
As expected, the likelihood-ratio technique
provides more reliable matches ($\approx7$ more sources using
our approach). 
Between the 347 counterparts
from the error-circle method and the 344 counterparts from the 
likelihood-ratio method (if a source has multiple counterparts, we chose
the one with the highest reliability parameter), 11 are different matches,
six of which have a likelihood-ratio counterpart with magnitude of $R>25.5$.

\begin{figure}
\centerline{
\includegraphics[scale=0.5]{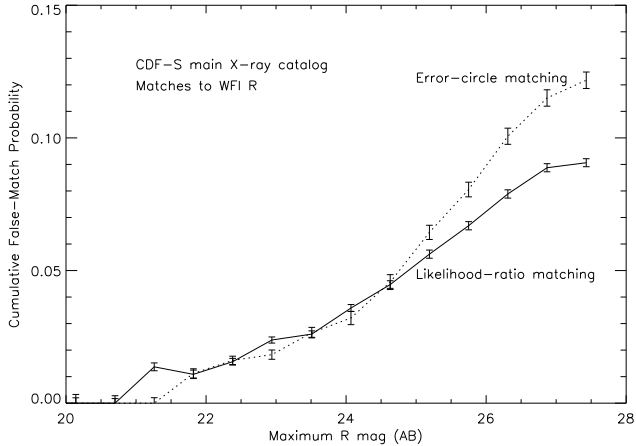}
}
\figcaption{
Cumulative false-match probability as a function of the 
\hbox{$R$-band} limiting magnitude
for the error-circle method (dotted curve) 
and the likelihood-ratio
method (solid curve). 
The false-match probabilities are estimated using the bootstrapping
technique described in \S\ref{results}. Error bars indicate the errors in the
mean values of the \hbox{false-match} 
probabilities at different magnitude limits.
The corresponding differential false-match probabilities
for the \hbox{likelihood-ratio} (error-circle)
method are
$\approx7\%$, $16\%$, $20\%$, and $24\%$ ($\approx7\%$, 21\%, 28\%, and 37\%) at
$R=24$, 25, 26, and 27, respectively.
The likelihood-ratio method yields more reliable matches at faint magnitudes.
The error-circle method appears to have a generally smaller false-match 
probability at bright magnitudes ($R\la23$); however, it also 
gives a smaller number of matches compared to the likelihood-ratio method,
and the numbers of expected true matches from these two methods are 
comparable.
\label{frwfi}}
\end{figure}

We studied the relation between the 
false-match probability and catalog depth by setting different 
$R$-band magnitude limits. The results are shown in Figure~\ref{frwfi},
which indicate that the likelihood-ratio method outperforms the 
error-circle method at faint magnitudes, and this is consistent with
the expectation
that the likelihood-ratio method can guard against false matches to faint
sources. Indeed, the 
six different counterparts with $R>25.5$ derived from the two methods 
contribute to 20\% of the total population of 30 $R>25.5$ counterparts for the
entire X-ray sample \citep[see also][]{Brusa2009}.

From the cumulative distributions of false-match probabilities
shown in Figure~\ref{frwfi}, we calculated the differential values 
at several $R$-band magnitudes. 
For the likelihood-ratio (error-circle) 
method, the \hbox{false-match} probabilities are
$\approx7\%$, $16\%$, $20\%$, and $24\%$ ($\approx7\%$, 21\%, 28\%, and 37\%) at
$R=24$, 25, 26, and 27, respectively, showing the challenge of deriving
secure matches at faint optical magnitudes, 
even with the \hbox{likelihood-ratio} technique.

As a further test, we performed the matching with the 
deeper GOODS-S $z$-band catalog using the error-circle method.
There are 266 matches and a $20\%$ false-match probability, while 
the likelihood-ratio method yields 259 matches and a $9\%$
\hbox{false-match} probability. 14 matches are different from the 
error-circle results. Ten of these different matches have $z>25.5$, 
contributing to $\approx30\%$ of all the counterparts with $z>25.5$.
Note here that the matching results and the
false-match probability of the error-circle method depend on the
choice of the matching radius. Using the same matching radius as that
for the WFI \hbox{$R$-band} might not be optimal for the
GOODS-S $z$-band. It is laborious to determine the optimal matching 
radius for the error-circle method, i.e., the matching radius that provides
a large number of counterparts without introducing too many false 
matches.\footnote{We tried several other matching radii and none of them can provide better
results than the \hbox{likelihood-ratio} method. For example, a matching
radius of $1.1\sqrt{\Delta_{\rm X}^2+\sigma_{\rm o}^2}$
($1.3\sqrt{\Delta_{\rm X}^2+\sigma_{\rm o}^2}$) resulted in
252 (261) matches with a 13\% (16\%) false-match probability.}
On the other hand, the likelihood-ratio method can perform 
a similar optimization (in 
terms of the completeness and reliability parameters) 
automatically and more carefully, and is thus 
more robust.

We conclude that the likelihood-ratio technique can provide much more reliable
matches than the error-circle method at faint optical magnitudes (e.g., 
$R\ga25$). The resulting matches may differ by $\approx20\%$ in this regime,
where we expect to find some of the most interesting objects in deep X-ray
surveys, such as heavily obscured AGNs and AGNs at very high redshifts.
When matching to shallower catalogs with low source densities, 
the \hbox{likelihood-ratio} and error-circle methods will produce similar results with
competitive reliabilities; however, 
the multiple candidates and reliability parameters provided by the 
likelihood-ratio matching are still useful and convenient when considering 
multiwavelength identifications.

\section{PHOTOMETRIC REDSHIFTS} \label{zphot}

The CDF-S area has been targeted intensively by photometric and spectroscopic 
surveys, allowing us to determine robustly the 
photometric redshifts (hereafter, photo-z's) 
of the \hbox{X-ray} sources. Compared to previous studies of the 
photo-z's of the optical sources \citep[e.g.,][]{Wolf2004} 
or X-ray sources \citep[e.g.,][]{Zheng2004} in the \hbox{CDF-S} region,
our photo-z work here has the following features and advantages: 
(1) We are utilizing the \hbox{X-ray} sources in the current $\approx2$~Ms \chandra\
catalog, with the most reliable counterpart information available
derived using the likelihood-ratio matching method.
(2) The most recent UV--to--IR deep surveys provide an unprecedented 
multiwavelength data set for photo-z calculation; the {\it Spitzer}, 
GEMS {\it HST}, and {\it GALEX} UV data have been used for the first time
in the photo-z calculation for the \hbox{CDF-S} X-ray sources. 
As we are dealing with a relatively small sample of sources, we were
able to carry out source deblending carefully and obtain the best-ever
photometric data for them (see \S\ref{phodata}).
The likelihood-ratio method
is used when matching the primary counterpart positions (Columns~4 and 5 in 
Table~\ref{cat}) to the 
photometric catalogs, and the false-match probabilities are small
($\la4\%$). Details regarding the multiwavelength data are described below.
(3) A collection of up-to-date secure 
spectroscopic redshifts (hereafter spec-z's) allows better
calibration of the SED templates and a fair assessment of the photo-z
quality.

Of the 446 identified X-ray sources, 220 have secure \hbox{spec-z's.}
We collected secure \hbox{spec-z's} ($\ga 95\%$ confidence level, with 
several secure spectral features)
from the following 
catalogs: \citet{LeFevre2004},
 \citet{Szokoly2004}, \citet{Zheng2004},
 \citet{Mignoli2005},
 \citet{Ravikumar2007}, \citet{Vanzella2008}, \citet{Popesso2009},
and Silverman et al. (2010, in preparation).
A matching radius of 0.5\arcsec\ was used when matching the
primary counterparts to the redshift catalogs.\footnote{We cannot apply 
the likelihood-ratio method described in \S2.2 to match the 
primary counterparts to 
the optical sources with secure redshifts, as these sources are not 
from well-defined and flux-limited catalogs.} A bootstrapping test similar
to that in \S2.3 indicates a false-match probability of
$0.8\%$ for these spec-z's.
We refer to these 220
sources with secure spec-z's as the spectroscopic sample.
Of the other 226 sources, six are identified as stars (see details below in
\S\ref{fitproc}), and we refer to the remaining 220 sources as
the photometric
sample. Both the spectroscopic and photometric samples have 220 X-ray
sources, and their photo-z's have been computed.
According to the X-ray and optical property analysis in
\citet{Bauer2004} and Xue et al.
(2010, in
preparation), more
than $\approx75\%$ of these sources
are AGNs (see also \S3.3), while the others appear to be starburst
or normal galaxies.

\subsection{Multiwavelength Photometric Data} \label{phodata}

We constructed a photometric catalog for the 446 identified X-ray
sources
including up to 42 bands of UV--to--IR data 
(up to 35 bands are used in the photo-z calculation 
for a given source; see below), derived from a number of publicly available 
catalogs.
For a source undetected in a given band, if the non-detection
is not caused by the possible blending with
a nearby source, we
estimated its 1~$\sigma$ magnitude upper limit, using
the background rms maps generated by {\sc SExtractor} \citep{Bertin1996}. 
The 1~$\sigma$ instead of the more conservative 3~$\sigma$
upper limits were calculated because these upper limits are used later 
in the photo-z calculation (see \S\ref{fitproc}).
Upper limits are not available for the COMBO-17 and MUSIC bands,
and we calculated only the $z$-band upper limits among the {\it HST} ACS bands.
We prioritized the catalogs as an indication of our preference for using
the data from these catalogs, with a highest priority of 5 and lowest 1.
The priorities are determined considering the spatial 
resolution and number of bands available (for the
consistency in colors, e.g., the MUSIC $U$-band has a higher priority
than the VIMOS $U$-band).
Details on the catalogs are given below.

\begin{enumerate}
\item
The {\it GALEX} UV catalog from {\it GALEX} Data Release 4, which includes the 
near-UV (NUV, $\lambda_{\rm eff}=2267$~\AA) and far-UV 
(FUV, $\lambda_{\rm eff}=1516$~\AA) data.\footnote{\url{http://galex.stsci.edu/GR4/}\,.}
The 5 $\sigma$ limiting
AB magnitudes for the NUV and FUV bands are 26.0 and 26.4, respectively.
This catalog has a priority of 1.

\item
The GOODS-S deep VIMOS $U$-band catalog \citep{Nonino2009},
with a 1 $\sigma$ limiting AB magnitude of
$29.8$.
This catalog has a priority of 1.

\item
The COMBO-17 optical catalog ($U$, $B$, $V$, $R$, $I$, plus 12 narrow-band 
filters; \citealt{Wolf2004,Wolf2008}). 
The 10 $\sigma$ limiting
AB magnitudes for the  $U$, $B$, $V$, $R$, and $I$ bands are
24.7, 25.0, 24.5, 25.6, and 23.0, respectively.
This catalog has a priority of 3.

\item
The MUSYC $BVR$-detected optical catalog ($U$, $B$, $o3$, $V$, $R$, $I$,
and $z$ filters; \citealt{Gawiser2006}),
and the MUSYC $K$-detected optical/NIR catalog ($B$, $V$, $R$, $I$, $z$, $J$,
and $K$ filters; \citealt{Taylor2009}). 
The 5 $\sigma$ limiting
AB magnitudes for the  $J$ and $K$ bands are 
23.1 and 22.4, respectively.
These catalogs have priorities of 2.

\item
The GOODS-S MUSIC optical/NIR/IR catalog ($U_{35}$, $U_{38}$, $U$,
$F435W$, $F606W$, $F775W$, $F850LP$, $J$, $H$, $K$, and IRAC filters;
\citealt{Grazian2006}). 
The 90\% limiting
AB magnitudes for the  $J$, $H$, and $K$ bands are 
24.5, 24.3, and 23.8, respectively
The $U$-band filter is the same as that used in
the GOODS-S deep VIMOS $U$-band catalog. 
This catalog has a priority of 3.

\item
The WFI $R$-band catalog (see \S2 of \citealt{Giavalisco2004}),
with a 5 $\sigma$ limiting AB magnitude of
$27.3$.
The $R$-band filter is the same as that used in
the COMBO-17 catalog.
This catalog has a priority of 2.

\item
The GOODS-S {\it HST} ACS $B$, $V$, $i$, and $z$-band 
($F435W$, $F606W$, $F775W$, and $F850LP$ filters) optical catalog,
with 5 $\sigma$ limiting
AB magnitudes of 29.1, 29.1, 28.5 and 28.2, respectively.
This catalog has a priority of 5.

\item
The GEMS $V$ and $z$-band ($F606W$ and $F850LP$) catalog 
\citep{Caldwell2008},\footnote{See \url{http://www.mpia-hd.mpg.de/GEMS/gems.htm} for 
the $V$-band data.} with 5 $\sigma$ limiting
AB magnitudes of 28.5 and 27.3, respectively.
This catalog has a priority of 4.

\item
The SIMPLE IRAC $3.6~\mu$m, $4.5~\mu$m, $5.8~\mu$m, and $8.0~\mu$m 
IR catalog \citep{Damen2010}, with 5 $\sigma$ limiting
AB magnitudes of 23.8, 23.7, 22.0, and 21.8, respectively. 
This catalog has a priority of 1.

\end{enumerate}

To collect photometric data from these catalogs for the 
X-ray sources,
we matched the primary counterparts of the X-ray sources to the UV, optical/NIR,
and IR catalogs using the same likelihood-ratio method described in 
\S\ref{method}. The bootstrapping exercises gave small false-match
probabilities ($\la4\%$).
The aperture-corrected total magnitudes were extracted from the 
catalogs for each matched source. 
Figure~\ref{sedexp} shows the SED of source 312 as an example of the resulting
multiwavelength photometric data.

Not all the data are used in the photo-z calculation.
We ignored the data if the errors are larger
than one magnitude.
When detections in the same filter
are present in different catalogs, we chose the data from the catalog
with the highest priority.
When a source has both MUSYC and COMBO-17 detections, the MUSYC optical
data (except for the $o3$ and $z$ bands) were ignored, as there are
a larger number of filters in the COMBO-17 data.
When a source has both MUSYC and MUSIC NIR detections, we neglected the 
MUSYC NIR data. Therefore, any individual 
source can have up to 35 filters of photometric
data that are used for deriving the photo-z.
We prefer the IRAC data
in the GOODS-S MUSIC catalog because the IRAC measurements are less affected by 
possible source-blending problems.
In cases when photometric data in two similar filters (e.g.,
GOODS-S MUSIC $U$-band and MUSYC $U$-band) differ by more than 0.5 magnitudes
(which may be caused by source variability or blending),
we used only the data with the higher priority.
For about ten GEMS sources, the $z$ and $V$-band magnitudes are $\ga1$
magnitude fainter
than the MUSYC $z$ and \hbox{$V$-band} data. We derived 
aperture photometry (with a 0.7\arcsec\ aperture radius) 
results using the GEMS science images, and they 
agree with the MUSYC photometry data. These sources appear to have some
low-surface-brightness emission, and the GEMS catalog underestimates 
their brightness. Therefore we ignored the GEMS data for these sources.
Upper-limit information was only used when there is no detection in a given
band and all other similar bands.

As we already know the presence of the X-ray sources and their primary
counterparts, we can search for their detections in the optical/NIR images
at low significance levels with less worry about false detections. 
Therefore, we ran {\sc SExtractor} on the WFI $R$-band and MUSYC optical/NIR
images with various detection significances down to 1~$\sigma$;
we did not search the GOODS-S and GEMS optical images
because
the corresponding catalogs are already very complete.
On average, only a few additional detections
were found in each band, and we visually examined the images to verify their
validity. 

To reduce the effects of source blending,
we inspected each source and corrected its 
photometric data if these are affected by blending.
About 10\% of the sources in the IRAC images and a few percent in the 
optical/NIR are blended with adjacent sources, and they are usually resolved
in the deep {\it HST} images, i.e., in the GOODS-S or GEMS $z$ band. 
We deblended these sources
by calculating the flux ratios using the pixel values at the source positions.
In some cases when there is only mild blending in the IRAC images or the 
blended sources are not detected in the $z$ band, we 
manually extracted the aperture fluxes by centering on the correct source 
positions (positions of the primary counterparts) 
and using an aperture radius of 1.5\arcsec; the fluxes were then
aperture corrected to give the total flux. 
For the one case where the primary counterpart is blended (source 22), we 
centered on the X-ray position to derive the IRAC fluxes.
We ignored the UV data
if the source suffers from blending, because the UV image resolution is not
as good as those for the optical/NIR/IR images. 

\begin{figure}
\centerline{
\includegraphics[scale=0.5]{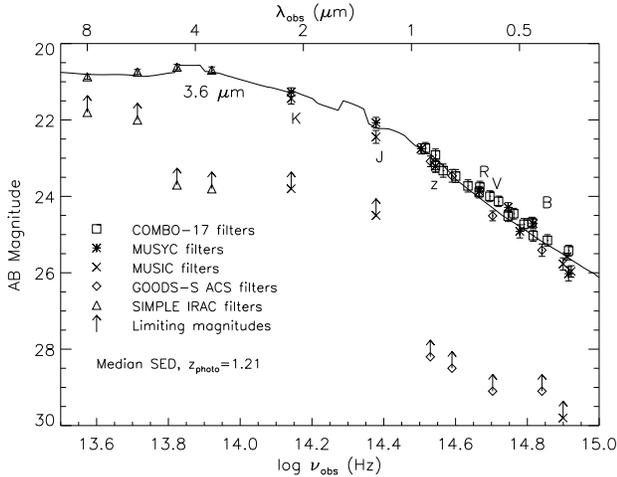}
}
\figcaption{
The observed median SED for the CDF-S X-ray sources and its best-fit SED template (a
hybrid type 2 AGN template).
Data in the same filter are combined (e.g., GOODS-S $z$ and GEMS $z$, MUSIC $U$
and VIMOS $U$). The NUV, FUV, and MUSIC $H$ bands are not shown as
the median magnitudes were calculated using only the detected sources
in these bands, and thus
the SED consists of 39 bands of data.
The MUSIC $J$, $K$, and
GOODS-S $B$ and $i$-band data only represent X-ray sources
in the GOODS-S region. The errors for the
median magnitudes are estimated using $NMAD/\sqrt{N_{med}}$.
The limiting magnitudes for some of the deep optical--to--IR bands are also
shown.
This fit should
not be interpreted physically as sources across a broad range of redshift
are included.
\label{sedmedian}}
\end{figure}

The final photometric catalog is presented in Table~\ref{phocat}.\footnote{A
FITS file containing Table~\ref{phocat} (without the first row) 
is available at 
\url{http://www.astro.psu.edu/~niel/cdfs/cdfs-chandra.html}.}
Rows~2--463 of Table~\ref{phocat} list the 
source \hbox{X-ray} IDs (Column~1) as well as
the photometric data, data errors, and data flags in each band:
the {\it GALEX} bands (Columns~2--7), VIMOS $U$ band (Columns~8--10), 
\hbox{COMBO-17} bands (Columns~\hbox{11--61}),
MUSYC bands (Columns~62--103), MUSIC bands (Columns \hbox{104--145}),
WFI $R$ band (Columns~\hbox{146--148}),
GOODS-S ACS bands (Columns \hbox{149-160}), GEMS ACS bands 
(Columns~\hbox{161-166}), SIMPLE IRAC bands (Columns~167--178),
and VLA 1.4 GHz band (Columns~179--181).
The radio data are used only in the X-ray source identification process, and
are not used in the photo-z calculation.
All the
photometric data are given in AB magnitudes, and are set to ``99''
if not available; upper limits are indicated by the ``$-$'' sign. 
A value of ``1'' for the data flag indicates that the data point was used
in the photo-z calculation, ``0'' means that the data point
was not used or not available, and ``$-$1'' means that the data point is 
probably problematic, either being blended or disagreeing with other data.
Row~1 of Table~\ref{phocat} provides, for the identified extragalactic
sources, the median magnitude (in the magnitude 
column), normalized median absolute
deviation ($NMAD$; in the magnitude-error column), and number of 
X-ray sources used to derive the median magnitude 
($N_{\rm med}$; in the data flag column) for each band.\footnote{The normalized
median absolute deviation is a robust measure of
the spread of the magnitudes ($\sigma$) in a given band, defined as
$NMAD=1.48\times {\rm median}(|m({\rm AB})-{\rm median}(m({\rm AB}))|)$ \citep[e.g.,][]{Maronna2006}.}
The median magnitude in a given band 
was calculated based on the magnitudes of
all the 440 sources (after
excluding the 16 unidentified sources and 6 stars), and for undetected 
sources, the limiting magnitude of that band was used.\footnote{The  
median magnitude and normalized 
median absolute deviation are stable parameters in 
general, and are not significantly affected by the choice of using the 
limiting magnitude or the 
individual magnitude upper limits for undetected
sources, as long as more than half
of the sources are detected.}
There are some exceptions: for the GOODS-S and MUSIC catalogs,
we used only X-ray sources within the coverage of these catalogs;
for the NUV, FUV, and VLA 1.4 GHz bands, in which less than
half of the X-ray sources are detected, and for the MUSIC $U$ and $H$ bands, 
which 
cover only a portion of the MUSIC field, we used only detected sources to
calculate the median magnitudes.
The data in Row~1 indicate how bright a 
typical CDF-S X-ray source is in each band; note that the median values in 
some bands do not sample the entire CDF-S X-ray source population.
The SED derived from the median magnitudes 
and its best-fit template are shown
in Figure~\ref{sedmedian} (see details about the SED template library and photo-z fitting procedure in \S~\ref{fitproc}). This fit should
not be interpreted physically as sources across a broad range of redshift
are included.

\begin{figure*}
\centerline{
\includegraphics[scale=0.5]{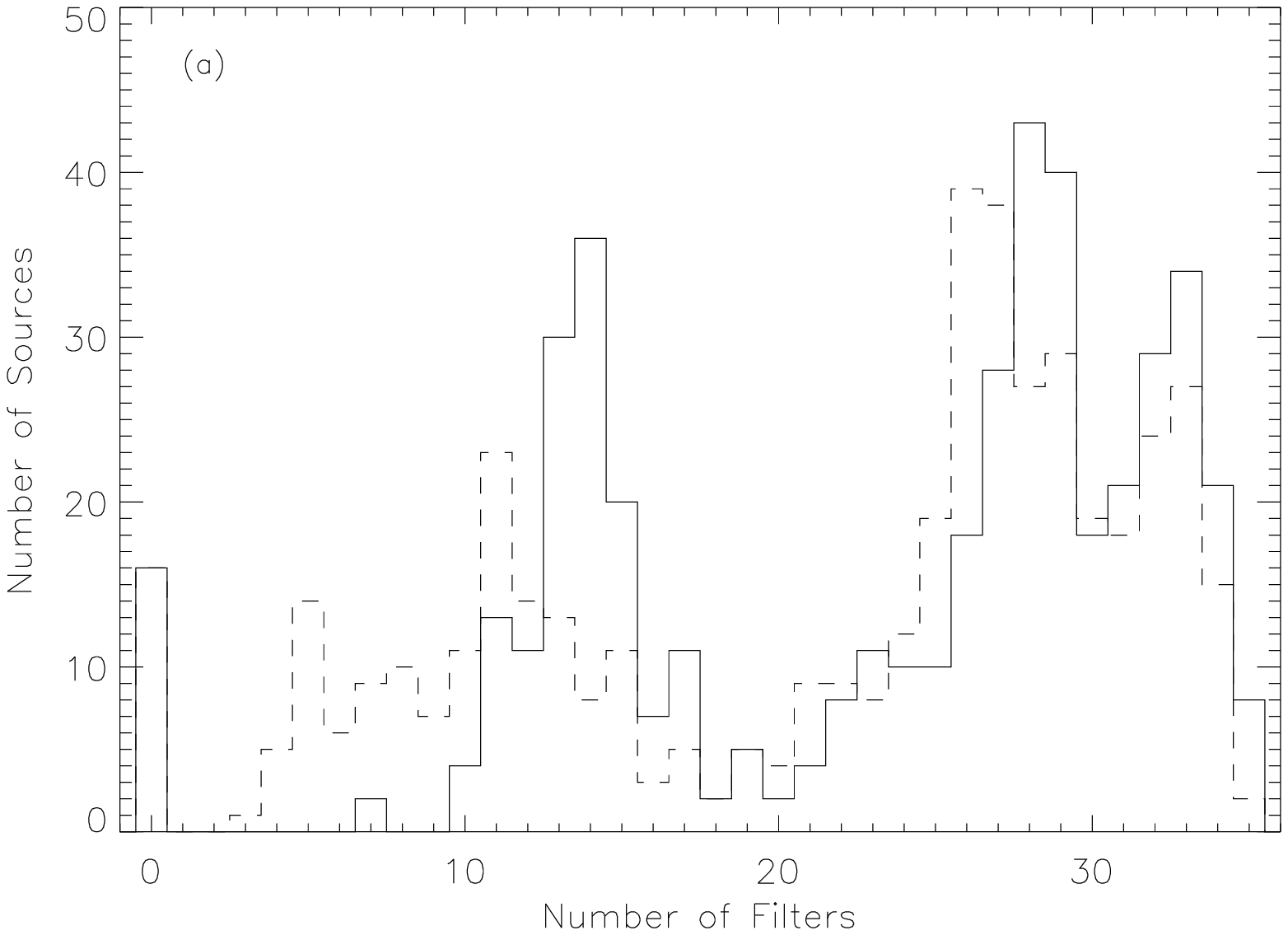}
\includegraphics[scale=0.5]{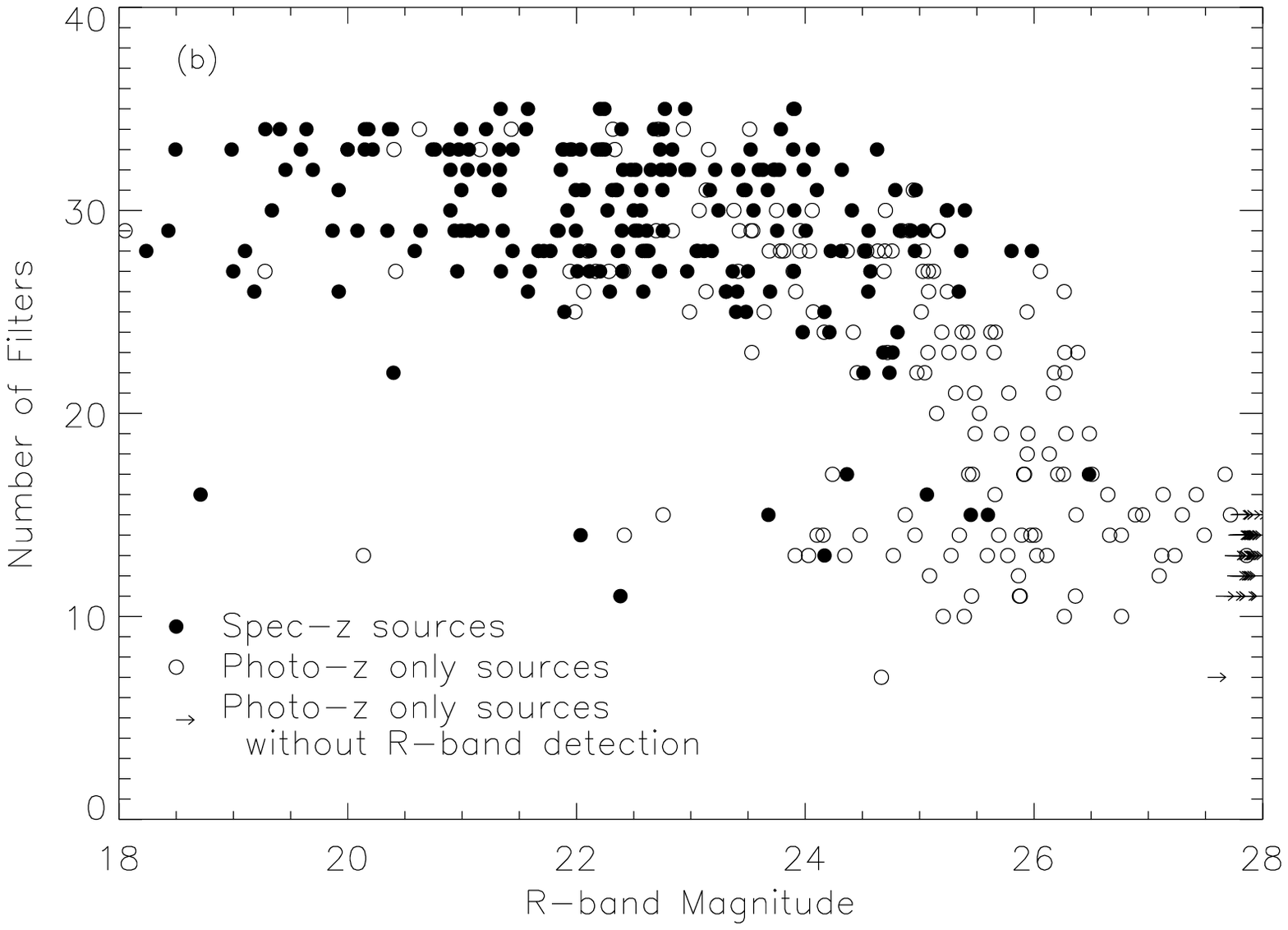}
}
\figcaption{
({\it a}) Distributions of the number of filters used in the photo-z 
calculation (solid
histogram) and the number of filters in which a source is detected (dashed
histogram). The difference between these two histograms
indicates the upper-limit
information used in the calculation.
The maximum number of filters used in the photo-z calculation
is 35, and the minimum number is 7.
The 16 sources without identifications have a filter number of 0.
The apparent bimodalities of the distributions are due to the inclusion or
exclusion of the COMBO-17 data, which gives a difference of $\approx17$ in the
number of filters.
({\it b})
Number of filters used in the photo-z calculation vs. the WFI/MUSYC $R$-band
magnitude for sources with spec-z's (filled dots) and sources without
spec-z's (open dots). For 62 sources that do not have \hbox{$R$-band}
data, their $R$-band upper limits are shown. Sources in the
spectroscopic sample are typically much brighter and have more
photometric data than the other sources. 
A few bright (e.g., $R<22$) sources have relatively small 
numbers of filters, because their COMBO-17 data are not available (not
detected due to source blending).
Also note that for very faint
sources, e.g., $R>26$, most of the data available are upper limits.
\label{filter}}
\end{figure*}

Figure~\ref{filter}a shows the distributions of the number 
of filters in which the source is detected,
and the number of filters used in the photo-z calculation (including upper
limit data).
The maximum values of these two filter numbers are both 35, and the minimum
values are 3 and 7, respectively.
We also show in Figure~\ref{filter}b that brighter $R$-band
sources tend to have
better multiwavelength coverage with a larger number of
filters.\footnote{For some sources that do not have a WFI $R$-band
detection, we used the MUSYC $R$-band magnitudes if available.
The MUSYC survey searched for sources
in the $B$, $V$, $R$, and $K$ bands and thus provides additional
detections.} Sources
in the spectroscopic sample are usually much brighter and have more
photometric data than the other sources.

\subsection{Photometric Redshift Fitting}\label{fitproc}

We use the Zurich Extragalactic Bayesian Redshift Analyzer (ZEBRA;
\citealt{Feldmann2006}) to compute the \hbox{photo-z's}.
ZEBRA provides a maximum-likelihood estimate for the 
photo-z's of individual galaxies or AGNs, combining together some useful
features such as 
automatic corrections of the SED templates to
best represent the SEDs of real sources at different redshifts (the training
procedure). 
Briefly, for every input SED template
in each user-specified redshift bin, ZEBRA 
generates a new SED template that 
simultaneously minimizes the $\chi^2$ values for all the training sources 
(with spec-z's available)
that use this input template as the best-fit template.
The procedure is iterated several times, selecting the \hbox{best-fit} template
from also the previous trained templates from the second iteration onward (see more
details in \citealt{Feldmann2006}).
The template correction/training process
substantially reduces template mismatches
and improves the photo-z accuracy.
In the template-training mode, we used the spectroscopic sample as a 
training set to improve the SED templates.

To construct a suitable library of SED templates, we followed a procedure
similar to that in \citet{Salvato2009}, which calculated photo-z's for the
{\it XMM-Newton} COSMOS sources. First, we ran ZEBRA on the spectroscopic sample of 220
sources, fixing the redshifts to their spec-z's and using the 259
PEGASE galaxy templates that cover different galaxy types (elliptical,
spiral, and starburst) and span a wide range of star-formation history
and intrinsic extinction \citep{Grazian2006}. Of these templates, 77 
were used by the 220 spectroscopic sources. Second, we repeated the above
procedure using the 10 AGN templates of \citet{Polletta2007} instead,
which include a variety of empirical quasar and Seyfert templates. 
Eight out of the 10 AGN
templates were used, including three type 1 and five type 2 AGNs.
Third, to account for contributions from both the nuclear emission 
and host-galaxy
emission in AGNs, we constructed a series of hybrid templates 
with the five most
frequently used galaxy templates and four most frequently used AGN
templates. 
For each pair of AGN and galaxy
templates, the templates were normalized by the total
3000--4000 \AA\ fluxes, and nine hybrids with varying galaxy/AGN ratios
(from 90:10 to 10:90) were created. This procedure produced a
set of 265 templates that consists of 180 hybrid templates, 77
galaxy templates, and 8 AGN templates. 
We chose a much larger sample of SED
templates than that in \citet{Salvato2009}, which has only 30 templates, 
so that we are less affected by selection effects when estimating
the photo-z quality using the spectroscopic sample (see \S\ref{zresults} below).
We have tested several smaller SED libraries, which provide 
little improvement in the \hbox{photo-z} quality for the spectroscopic sample, whereas
limited SED templates 
will probably hurt the performance of ZEBRA on the photometric sample as they
might not represent the typical SEDs of the sources without spec-z's.

When using ZEBRA to fit the SEDs, we set most of the parameters to
their default values.
The
filter responses and galaxy templates were resampled to a uniform wavelength 
mesh with a resolution of 4 \AA\ between 400 \AA\ and 10 $\mu$m.
To minimize possible spurious boundary effects, we ran
the template-correction mode twice with two different sets of 
redshift bins and then
averaged the resulting improved templates to create a final SED library.
For the input photometric catalog, we employed a minimum magnitude error
of 0.06 for each filter, 
which effectively avoids cases where a few data points 
with unrealistically small errors have very large weights and thus
drive the overall fitting results.
We applied a luminosity prior of
$M_B\le-18$, which forces a lower limit on the absolute $B$-band magnitude. 
There are only two sources in the spectroscopic sample 
with fainter magnitudes. Applying such a prior leads to higher photometric
redshifts
for such faint sources, but tests show that it reduces the percentage of
catastrophic photo-z failures (outliers in a spec-z vs. photo-z plot) 
considerably.
ZEBRA does not take upper limits in the input data. To circumvent this problem,
we used a ``workaround'' by treating the 1~$\sigma$ upper limits as actual
detections and setting the flux ranges to be from 0 to 2~$\sigma$ 
(R. Feldmann 2009, private communication).
These data points generally have a small weight in the SED fitting
because of the large error bars, but they do help ZEBRA to determine the correct
location of the Lyman break and thus give a better estimate of the photo-z,
especially for sources with upper limits only
in the NUV/FUV, and sources with detections only in the NIR/IR.

For each source, the ZEBRA SED fitting procedure 
generates a two-dimensional probability 
distribution in template and redshift space; the probabilities indicate
the quality of fit based on the maximum-likelihood analysis. 
The best fit with the highest
probability (corresponding to the lowest $\chi^2$ value) 
is chosen as the final photo-z and SED template for each source.
In the photometric sample, $\approx55\%$ of the sources 
having multiple
local maximum probabilities.
As a supplement, we compare these local maximum probabilities,
select the fit with the second highest probability
(corresponding to the second lowest $\chi^2$ value),
and derive an alternative photo-z, if the 
second highest probability is within two orders of magnitude 
of the highest value ($\approx40\%$ of the sources 
with multiple
local maximum probabilities).
If available, alternative photo-z's are provided for the photometric sample 
in Table~\ref{zcat} (see below).

Stars in the sample were identified by visual inspection of 
the GOODS-S and GEMS {\it HST} images and also the SED fitting results
(the AGN/galaxy templates we used generally cannot fit a star SED and there is
a strong optical excess). Every source was visually examined, 
though we focused on 
sources identified as stars by the COMBO-17 and \hbox{GOODS-S} MUSIC 
classifications, and sources with low full-band \hbox{X-ray--to--$z$-band} 
flux ratios. There are six stars identified, with X-ray source
IDs 36, 172, 206, 352, 400, and 420; all of them have flux ratios 
($F_{\rm 0.5\textrm{--}8.0~keV}/F_z$) smaller than 0.01.

\subsection{Photometric Redshift Results} \label{zresults}

Photo-z's were obtained for the 440 X-ray sources in the
spectroscopic and photometric samples. 
They are listed in Table~\ref{zcat}, with the details of the columns 
given below.
\begin{itemize}
\item
Column~(1): the X-ray source ID number.

\item
Column~(2): the spec-z, set to ``$-$1'' if not available.

\item
Column~(3): the photo-z derived using ZEBRA. Set to ``0'' for the six
stars and ``$-1$'' for the 16 sources without identifications.

\item
Columns~(4) and (5): the 1~$\sigma$ confidence interval (lower and upper bounds)
of the photo-z.
We caution that the error estimate is probably not be a good indicator of the 
accuracy of the photo-z (see \S\ref{zaccu} for more details).
 
\item
Columns~(6) and (7): the number of filters in which the source is detected, 
and the number of filters used in the photo-z calculation. The 
difference between these two numbers is the number of upper limits used;
see Table~\ref{phocat} and Figure~\ref{filter} for more details.

\item
Column~(8): the reduced $\chi^2$ value for the best-fit result, 
$\chi^2_{\rm dof}$, indicating
the quality of fit. A value that deviates significantly from unity is generally
caused by disagreement between the photometric 
data in different catalogs, which is often likely
the result of AGN variability.

\item
Column~(9): the alternative photo-z, set to ``$-$1'' if not available. 
There are 49 sources having an alternative photo-z.

\item
Column~(10): the reduced $\chi^2$ value for the alternative photo-z,
$\chi^2_{\rm dof,alt}$.

\item
Column~(11): a flag indicating 
whether the X-ray source is detected in the optical.
A value of ``0'' means no optical detection, in which case the photo-z
was calculated using only the IR/NIR data and the optical 
upper-limit information (and is probably not very 
reliable). There are 20 sources in the photometric sample that do not 
have optical detections.

\item
Column~(12): Spec-z reference number; numbers 1 to 8 correspond to 
\citet{LeFevre2004},
 \citet{Szokoly2004}, \citet{Zheng2004},
 \citet{Mignoli2005},
 \citet{Ravikumar2007}, \citet{Vanzella2008}, \citet{Popesso2009},
and Silverman et al. (2010, in preparation), respectively.
\end{itemize}

\begin{figure*}
\centerline{
\includegraphics[scale=0.5]{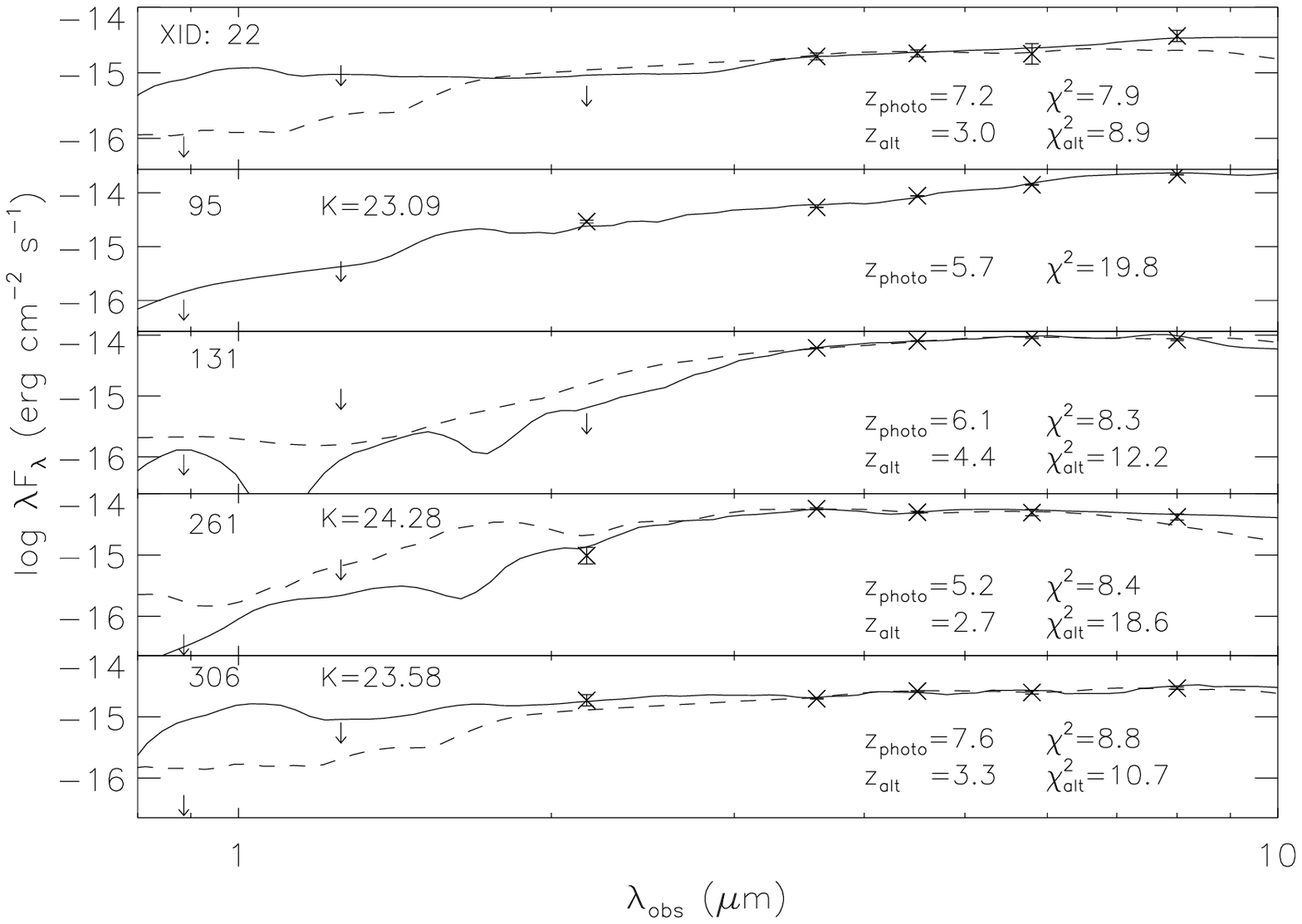}
\includegraphics[scale=0.5]{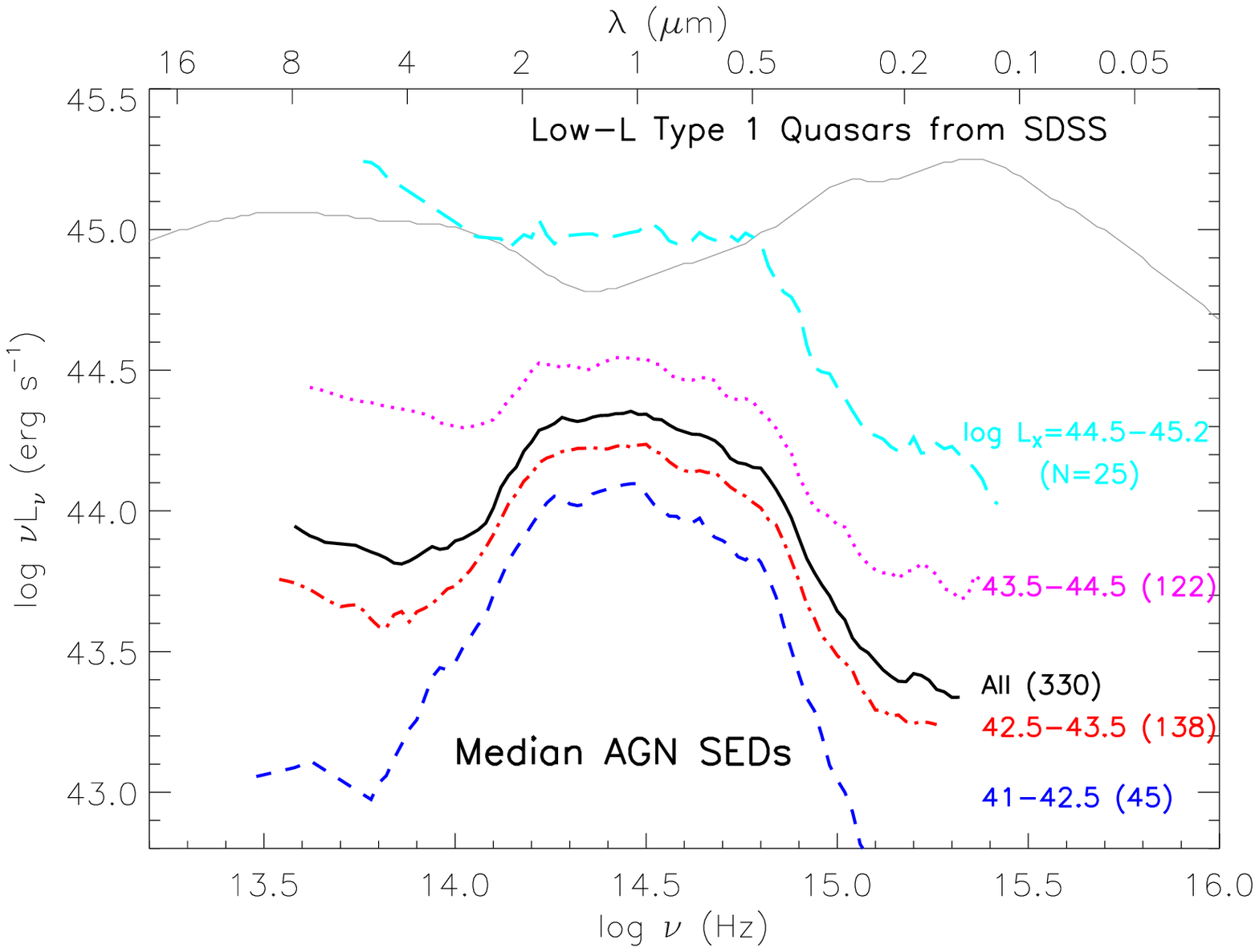}
}
\figcaption{
({\it a}) SEDs and best-fit templates (solid curves) 
for the five X-ray sources with \hbox{photo-z's}
greater than 5. Also shown are the alternative fitting templates (dashed 
curves) for four of the sources. 
These templates are either galaxy templates or hybrid AGN templates.
None of the five sources is detected blueward
of $K$ band (the $K$-band magnitude is 
indicated in each panel if available).
The 1~$\sigma$ upper limits on the $z$, $J$, and $K$
bands are shown if there are no detections. The 1~$\sigma$ upper limits
do not set tight constraints on the actual fluxes, 
so they may appear below the the SED templates
in some cases. ({\it b}) The median SEDs of $z=0$--4
AGNs in the CDF-S, derived from the 
photometric data and best-fit SED templates. 
The 24~$\mu$m data from the {\it Spitzer} Far Infrared Deep Extragalactic
Legacy Survey (FIDEL; Dickinson et~al. 2010, in preparation) 
were also used. The cyan long-dashed, magenta dotted, 
black solid, red dash-dotted, and blue dashed curves represent median SEDs
in different X-ray luminosity bins (as labeled, along with the number of
AGNs contributing to each median SED). 
The grey solid curve represents the 
low-luminosity quasar mean SED from \citet{Richards2006} with the host-galaxy
contribution subtracted. The SEDs of CDF-S AGNs are generally dominated
by host-galaxy light and show heavy extinction in the UV.
[{\it See the electronic
edition of the Journal for a color version of this figure.}]
\label{f09}}
\end{figure*}

The highest spec-z for a CDF-S X-ray source is 3.70. However,
the derived \hbox{photo-z's} show that ten sources in the photometric sample 
have \hbox{$z_{\rm photo}=4.3$--7.6}. These sources are generally not detected in the optical,
or only have a faint optical counterpart. They are detected in 4--10
photometric bands, and the resulting photo-z's are probably not very reliable.
In fact, eight of these objects have an alternative photo-z of 
\hbox{2.7--4.4} as listed in
Table~\ref{zcat} (for comparison, only 41 of 
the remaining 210 sources in
the photometric sample have an alternative photo-z).
In Figure~\ref{f09}a, we show the SEDs along with
the best and alternative fitting templates of the five sources with
$z_{\rm photo}>5$ (source 95 does not have an alternative photo-z). 
The SEDs and fitting templates for the five $z_{\rm photo}\approx4$--5 sources
(XIDs 7, 21, 191, 404, and 439) are qualitatively similar to those shown in Figure~\ref{f09}a.
Given the limited SED data available, both the best and alternative fitting 
results appear acceptable. 
We estimated the intrinsic hydrogen column
densities ($N_{\rm H}$; see the end of this section 
for details of $N_{\rm H}$ estimation)
for these sources.
Two sources (XIDs 7 and 439) have estimated $N_{\rm H}$ values smaller
than $6\times10^{22}$~cm$^{-2}$, and another seven
sources have $N_{\rm H}$ values between $1\times10^{23}$
and $8\times10^{23}$~cm$^{-2}$.
Source 261 has a very hard
X-ray spectrum, which requires a Compton-thick column
density ($N_{\rm H}\approx1.9\times10^{24}$~cm$^{-2}$; an outlier compared to
the $N_{\rm H}$ values of all the other X-ray sources) to produce the observed
X-ray counts at $z_{\rm photo}=5.2$, and is thus more likely located
at its alternative redshift of $z_{\rm alt}=2.7$, which only requires
$N_{\rm H}\approx8\times10^{23}$~cm$^{-2}$.
Given the properties above and 
the scarcity of $z>4$ AGNs in the \chandra\ deep fields, e.g., only three such
objects are identified spectroscopically in the $\approx2$ Ms \hbox{CDF-N} 
\citep[e.g.,][]{Waddington1999,Vignali2002,Barger2003}\footnote{One 
X-ray source (XID 618) in the \citet{Giacconi2002} catalog for the $\approx1$ 
Ms CDF-S has a spec-z of 4.759 \citep{Vanzella2006,Fontanot2007}. The source
was detected only by {\sc SExtractor} in \citet{Giacconi2002}, and was not
detected in the \citet{Alexander2003} catalog for the $\approx1$
Ms CDF-S or the L08 catalog for the $\approx2$
Ms CDF-S.}, 
we consider that at least half of the ten sources with 
high \hbox{photo-z's} are probably
at intermediate redshifts and the alternative \hbox{photo-z's} are
preferable for them. 
In the following analysis, we
will still use the best-fit photo-z's for all the sources.

\citet{Rodighiero2007} estimated photo-z's for a sample of 
optically undetected
objects, including three of our X-ray sources (XIDs 289, 306, and 322, with
photo-z's $\approx4$, $\approx4$, and $\approx8$). 
Their estimated photo-z for source 306 is more 
consistent with its alternative photo-z in our catalog 
(see Fig.~\ref{f09}a).
We note that \citet{Mainieri2005} also derived a single solution for the
photo-z of source 95 (see Fig.~\ref{f09}a), 
though with a higher value, $z_{\rm photo}=7.9$;
this source may be a $z>5$ AGN candidate. 
There are also at least two $z>5$ AGN candidates among the unidentified 
X-ray sources (see \S\ref{noiddis}).
\citet{Koekemoer2004} studied seven extreme X-ray--to--optical ratio sources
(EXOs) in the $\approx1$~Ms CDF-S. These sources are all included in the
L08 catalog, with source IDs of 95, 98, 133, 189, 235, 290,
and 404. Sources 95 and 404 have photo-z's of 5.7 and 4.7 in our catalog, 
respectively,
while the other five sources have photo-z's ranging from 2.2 to 3.2.
It thus appears that most of the EXOs are at moderate redshifts 
\citep[e.g.,][]{Mainieri2005}.

\begin{figure*}
\centerline{
\includegraphics[scale=0.5]{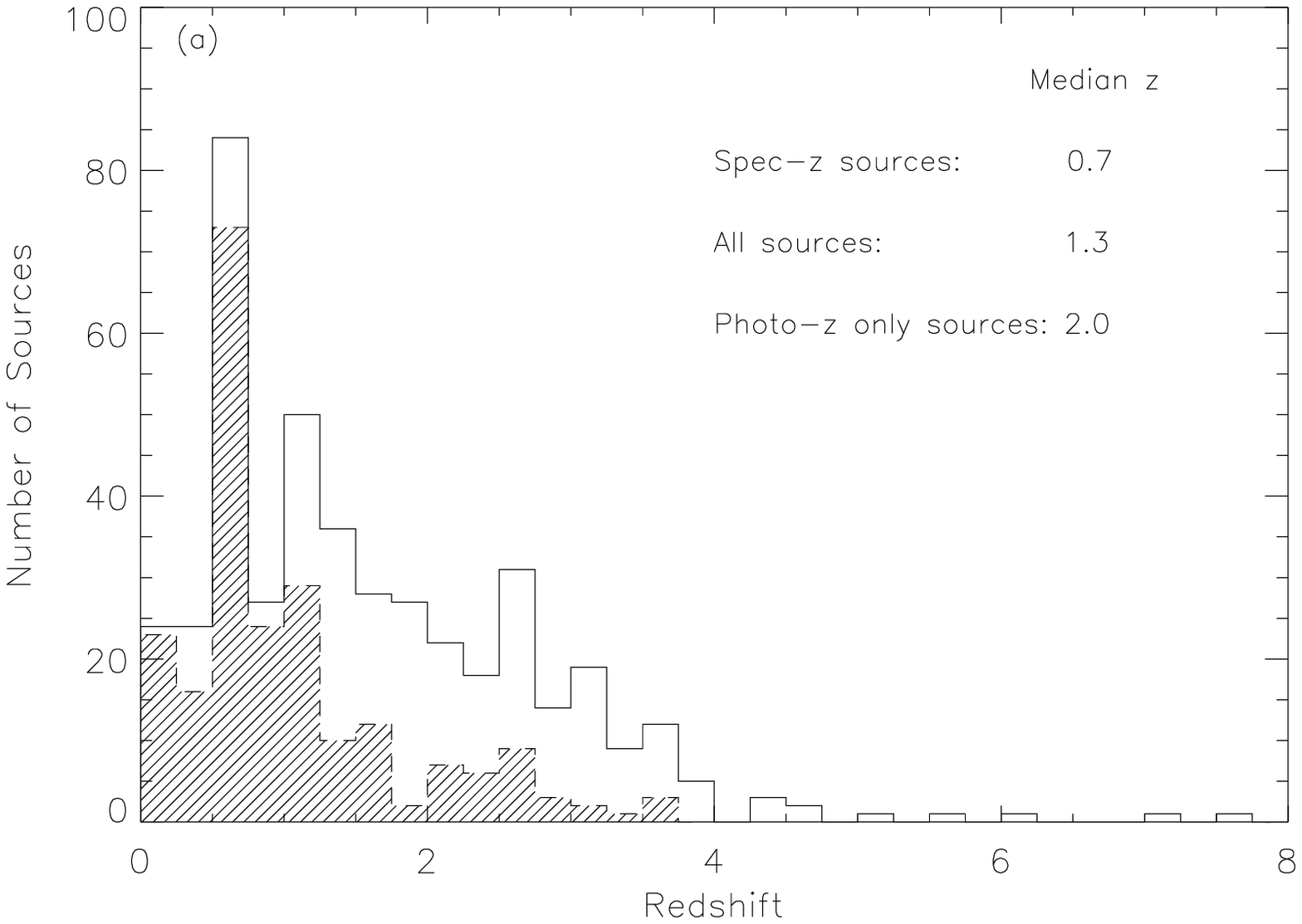}
\includegraphics[scale=0.5]{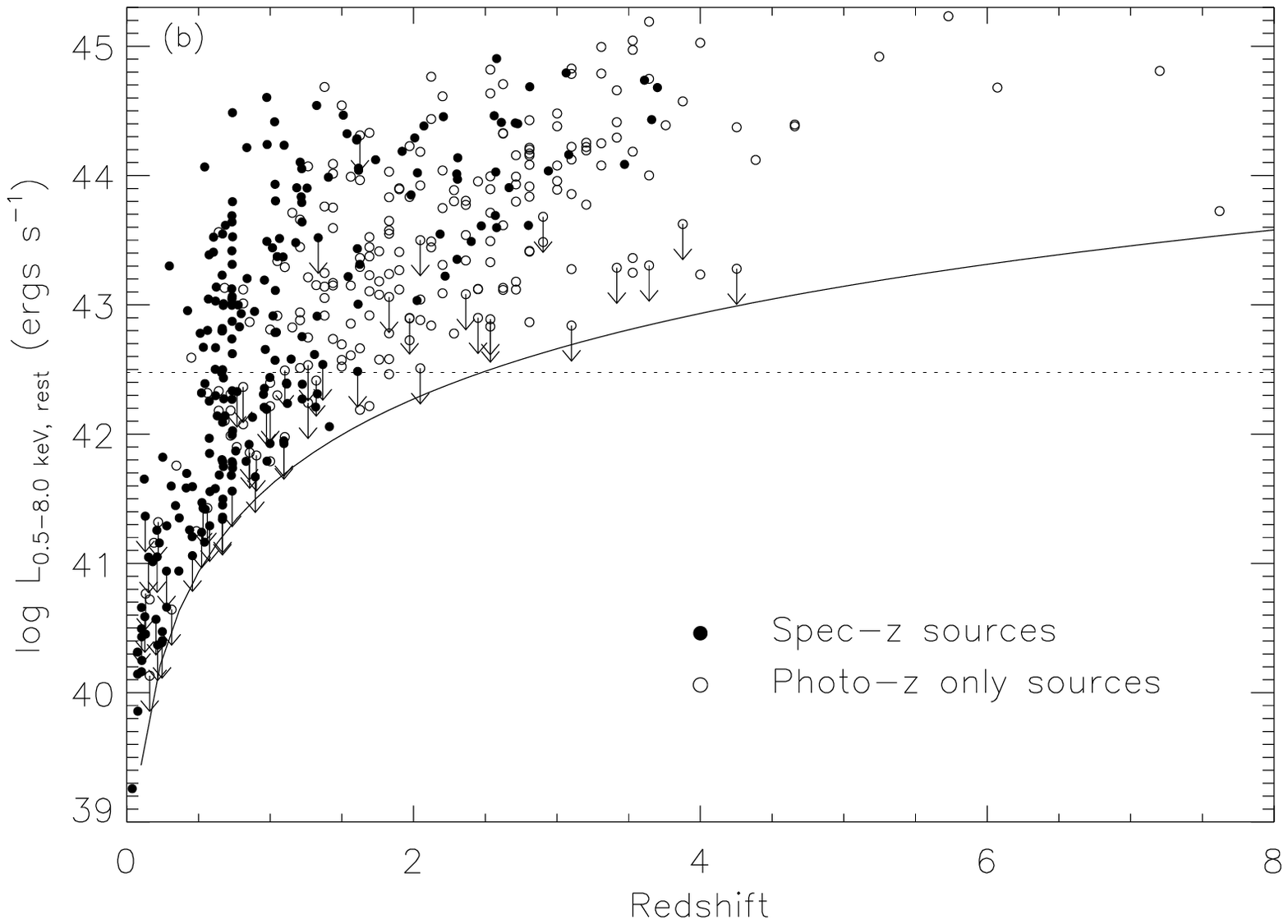}
}
\figcaption{
({\it a}) Histogram showing the distribution of photo-z's. The shaded region
shows the distribution of spec-z's. Sources with only photo-z's are
generally at higher redshifts than spec-z sources. ({\it b})
Redshift-luminosity distribution for spec-z sources (filled dots)
and photo-z sources (open dots). The solid curve represents
the limiting
0.5--8.0 keV luminosity at the center of the CDF-S region (L08), assuming
$\Gamma=1.8$ and $N_{\rm H}=10^{22}$~cm$^{-2}$.
The dotted line indicates
$L_{\rm 0.5\textrm{--}8.0~keV,rest}=3\times10^{42}$~\lum. 
Some of the photo-z's for the ten objects 
with $z_{\rm photo}>4$ are probably not
reliable.
\label{zall}}
\end{figure*}

The distribution of all the derived photo-z's is shown in Figure
\ref{zall}a. The median photo-z for all the sources is 1.3, and it is 2.0
for just the photometric sample. For comparison, the median spec-z
is only 0.7 for the spectroscopic sample.\footnote{This median
spectroscopic redshift may be partially related to the 
two apparent redshift spikes at $z=0.67$ and 0.73 in the CDF-S 
\citep[e.g.,][]{Gilli2003,Silverman2008}.} 
Between $z=1$ and $z=4$, there are 179 sources in the photometric sample 
and only 84 sources in the spectroscopic sample.
It is reasonable that
sources without spectroscopic information are in general at higher
redshifts. We calculated the rest-frame 0.5--8.0 keV luminosities
for the \hbox{X-ray} sources using the \hbox{spec-z's} and photo-z's.
The intrinsic absorption was corrected for by assuming an intrinsic X-ray
power-law photon index of $\Gamma=1.8$ and using 
XSPEC (Version 12.5.0;
\citealt{Arnaud1996}) to derive the appropriate intrinsic hydrogen column 
density ($N_{\rm H}$) that
produces the observed \hbox{X-ray} band ratio (defined as the ratio of count rates
between the hard and soft bands).
A plot of
luminosity versus redshift is shown in Figure
\ref{zall}b. Sources in the photometric sample have larger luminosities than
spec-z sources on average.
We selected AGNs from the X-ray sources following the
criteria used by \citet{Bauer2004}. Sources with 0.5--8.0~keV 
\hbox{X-ray} luminosities
$L_{\rm 0.5\textrm{--}8.0~keV,rest}>3\times10^{42}$~\lum, 
effective photon indices $\Gamma_{\rm eff}<1$, or \hbox{X-ray--to--optical} 
($R$-band)
flux ratios $F_{\rm 0.5\textrm{--}8.0~keV}/F_R>0.1$ are considered to be AGNs.
In this way, we classified 354 AGNs among the 440 identified X-ray sources,
150 in the spectroscopic sample and 204 in the photometric sample. 
It thus appears that the majority of the X-ray sources in the photometric
sample are AGNs at relatively high redshifts. In Figure \ref{f09}b,
we show the median SED of 
330 AGNs with detections in at least six photometric bands 
and having redshifts $z<4$, along with the median SEDs in several different X-ray 
luminosity bins. The median SEDs in the optical are dominated by stellar light from 
the host galaxies, consistent with our SED fitting results (see \S3.6).

\subsection{Photometric Redshift Accuracy}\label{zaccu}

\begin{figure*}
\centerline{
\includegraphics[scale=0.5]{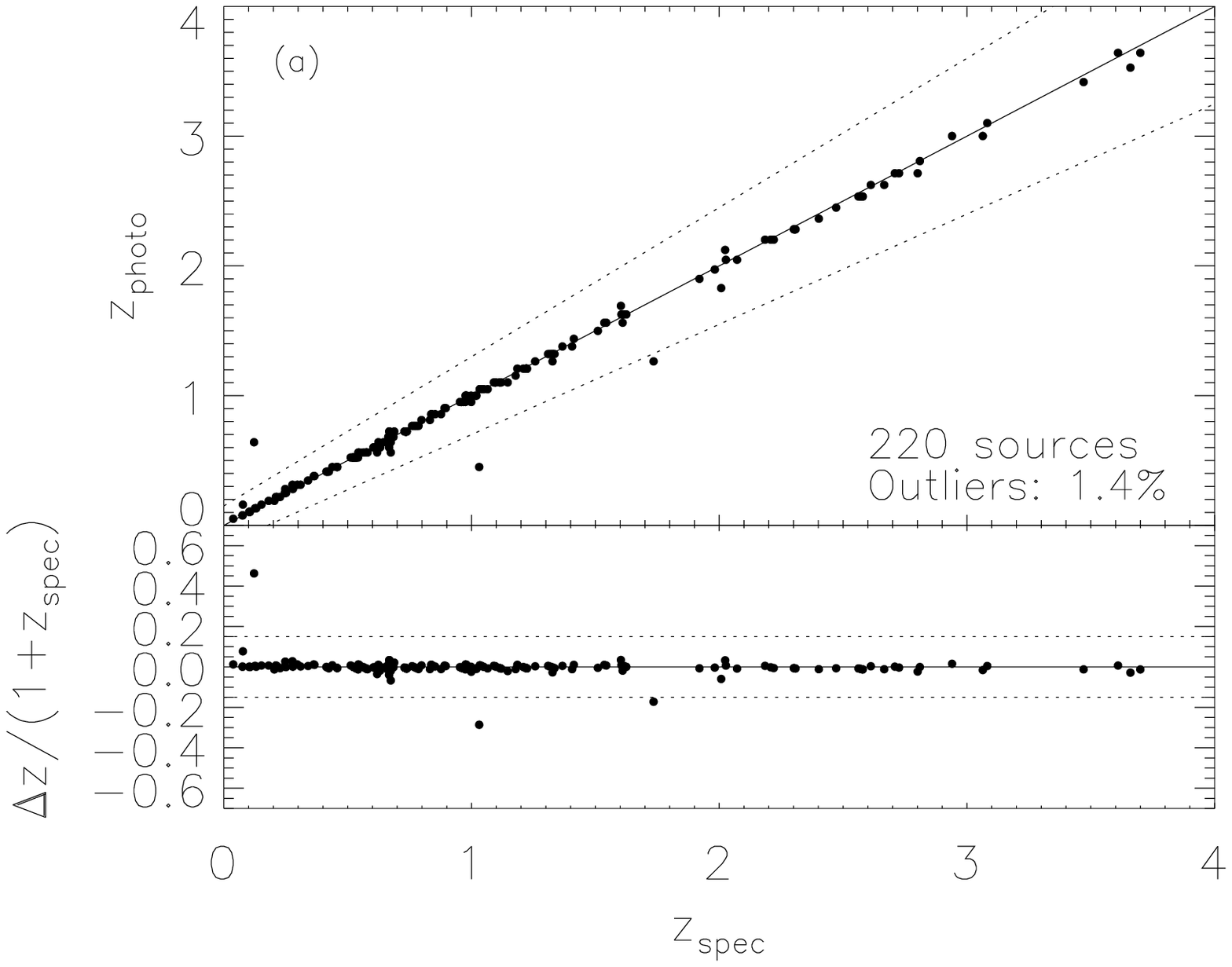}
\includegraphics[scale=0.5]{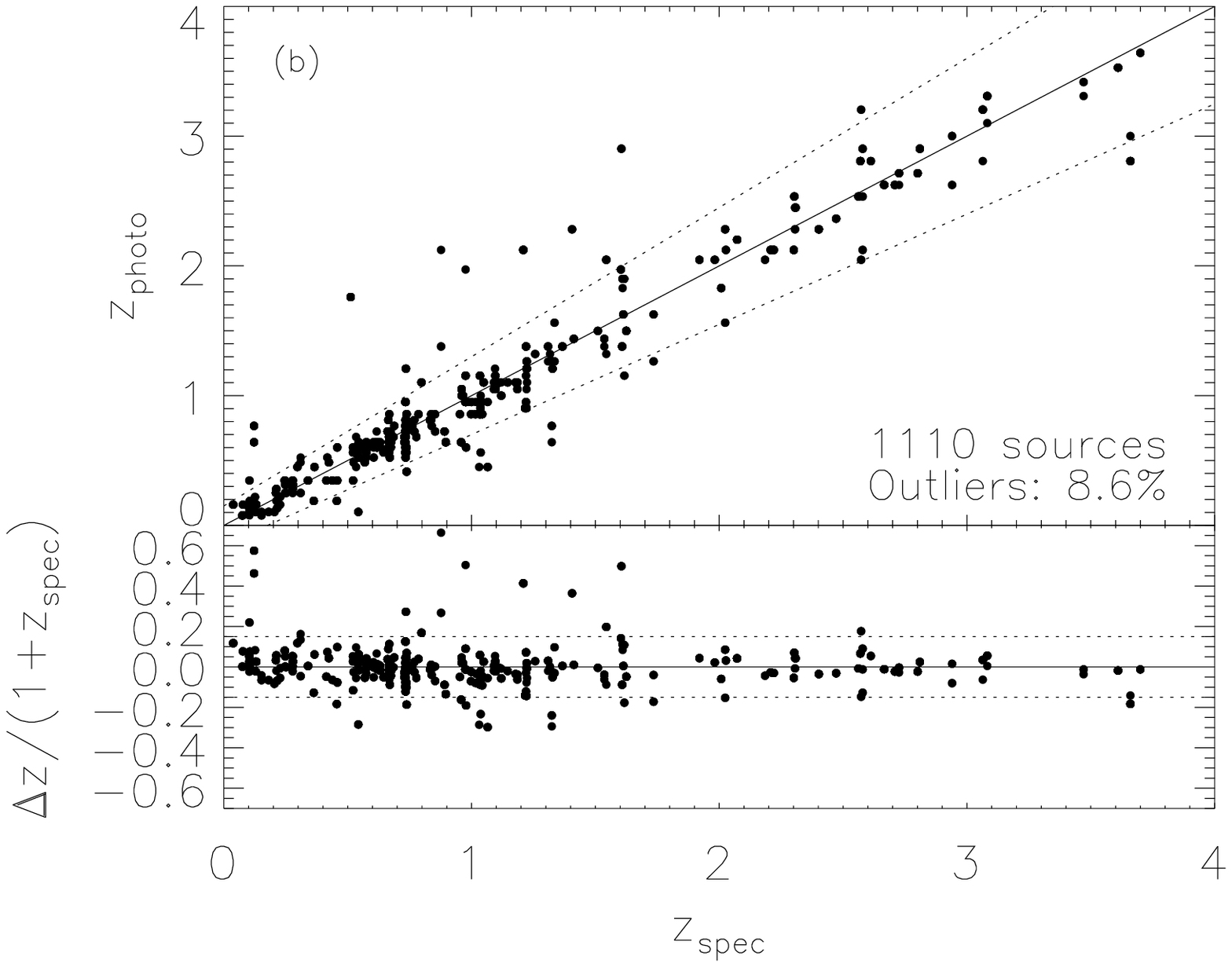}
}
\figcaption{
Comparison of the photo-z's and spec-z's for
({\it a}) the spectroscopic sample and ({\it b}) the blind-test sample. 
The upper panels show the distribution of $z_{\rm spec}$ vs. $z_{\rm photo}$,
and the lower panels show $z_{\rm spec}$ vs. $\Delta z/(1+z_{\rm spec})$.
The solid lines indicate $z_{\rm photo}=z_{\rm spec}$, and
the dotted lines represent relations of
$z_{\rm photo}=z_{\rm spec}\pm0.15(1+z_{\rm spec})$, which define the
outlier limits.
Note that
one data point can represent several sources in the blind-test sample ({\it b}).
The photo-z's and
spec-z's agree very well for the spectroscopic sample, while the \hbox{photo-z's} and
spec-z's differences are considerably larger for the blind-test sample.
The photo-z accuracy does not appear to degrade at high redshifts ($z\approx2$--4).
\label{zplot}}
\end{figure*}

\begin{figure*}
\centerline{
\includegraphics[scale=0.5]{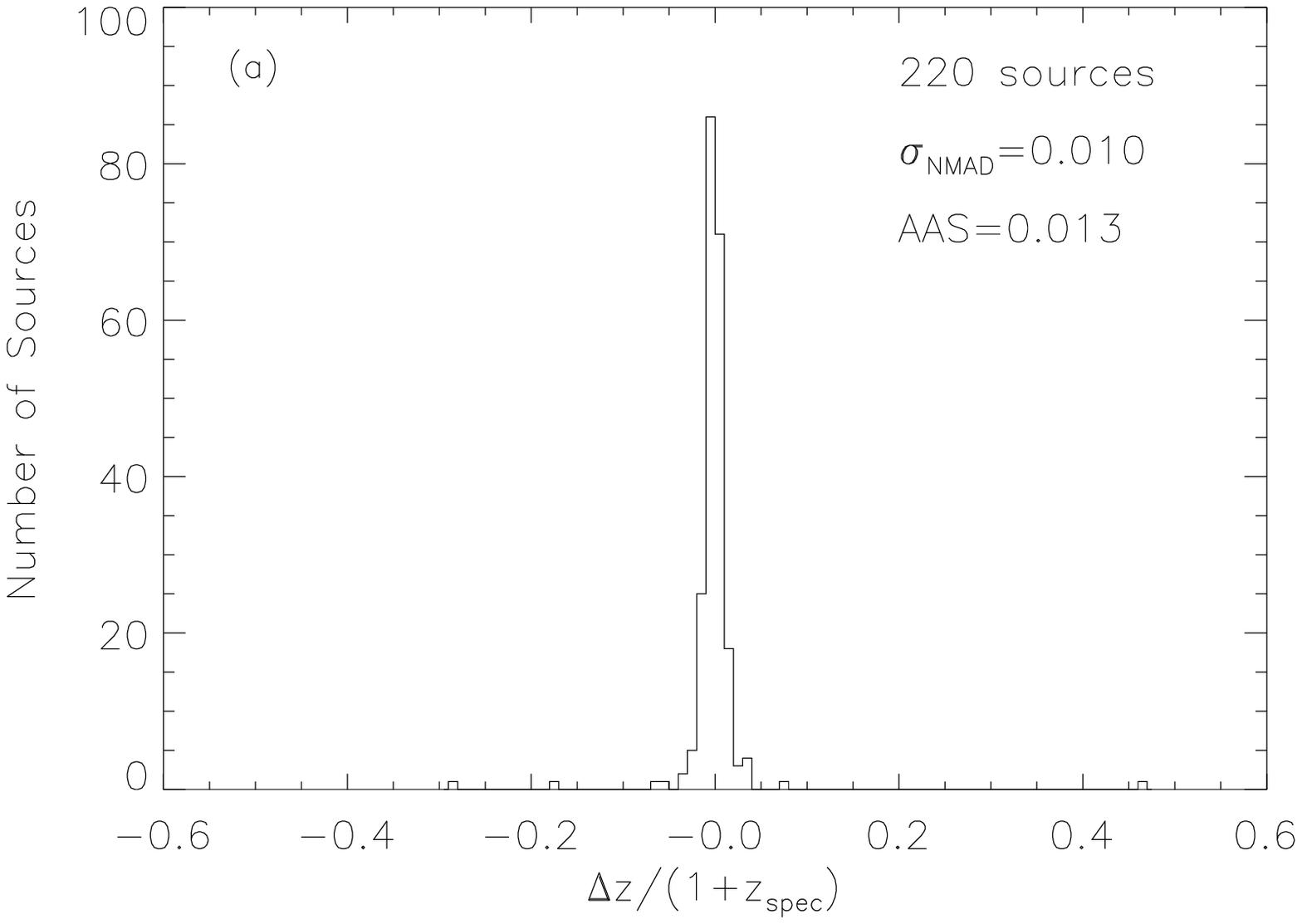}
\includegraphics[scale=0.5]{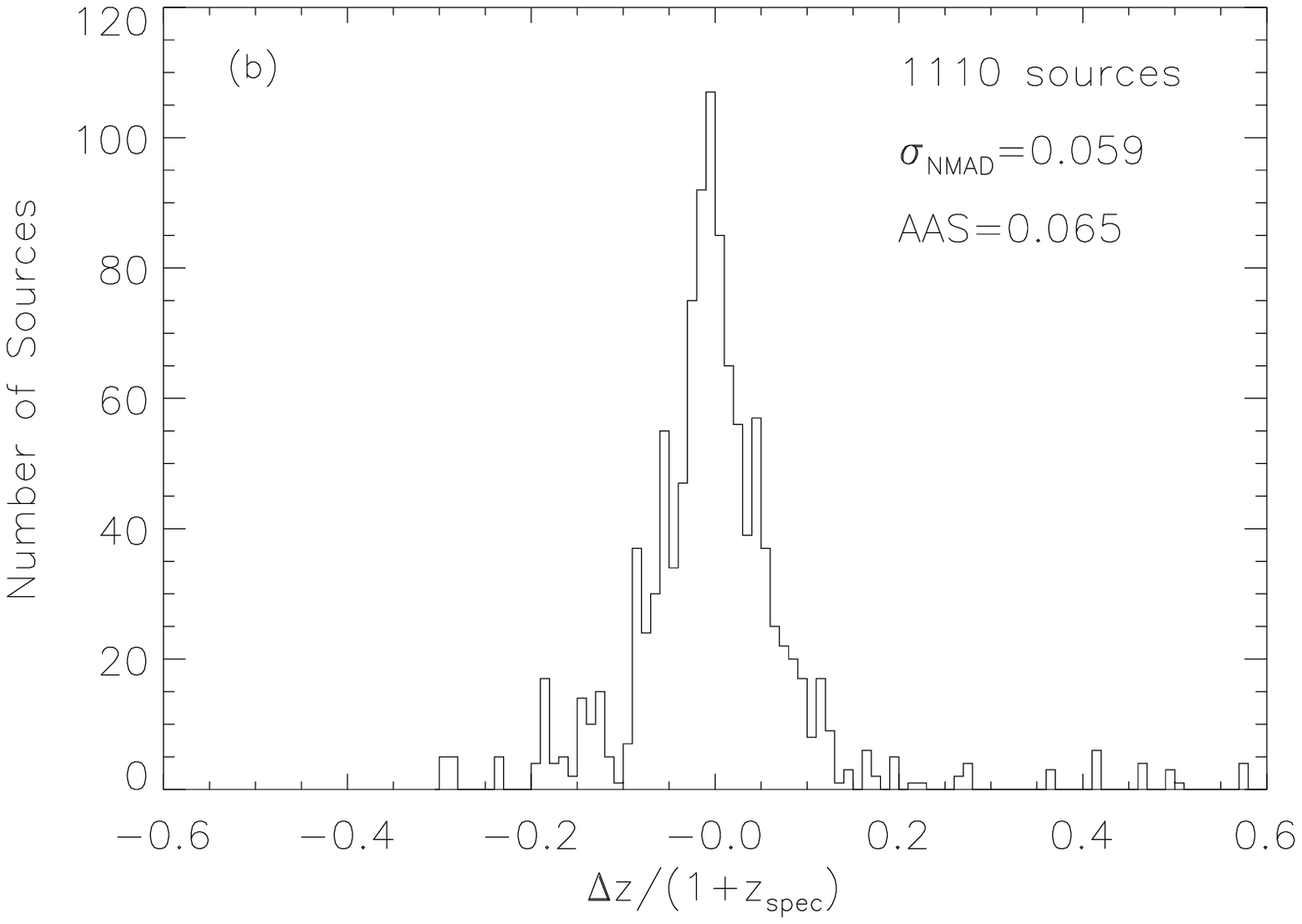}
}
\figcaption{
The distribution of the photo-z accuracy, $\Delta z/(1+z_{\rm spec})$,
for ({\it a}) the spectroscopic sample and ({\it b}) the blind-test sample.
\label{zhist}}
\end{figure*}

\begin{figure*}
\centerline{
\includegraphics[scale=0.5]{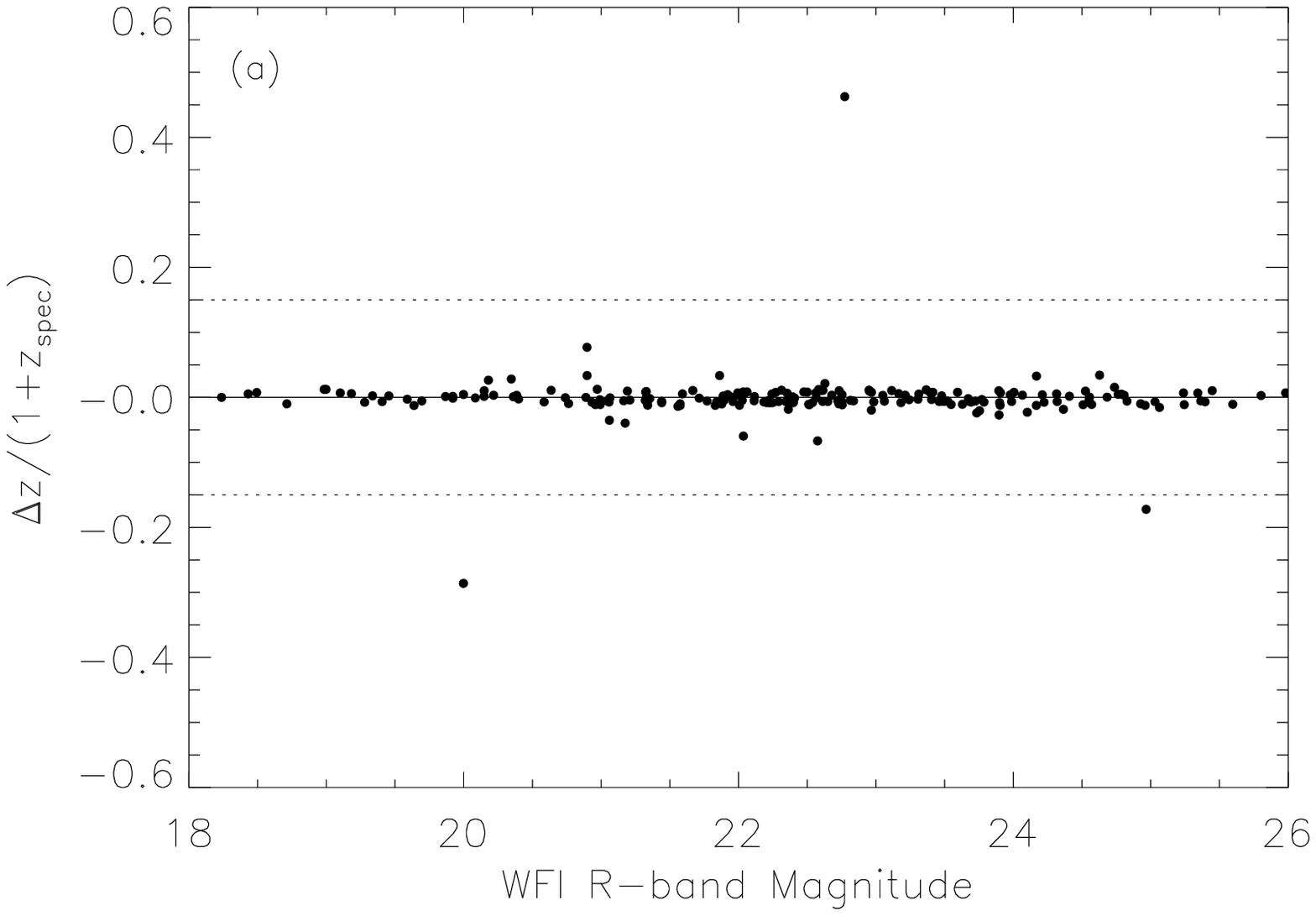}
\includegraphics[scale=0.5]{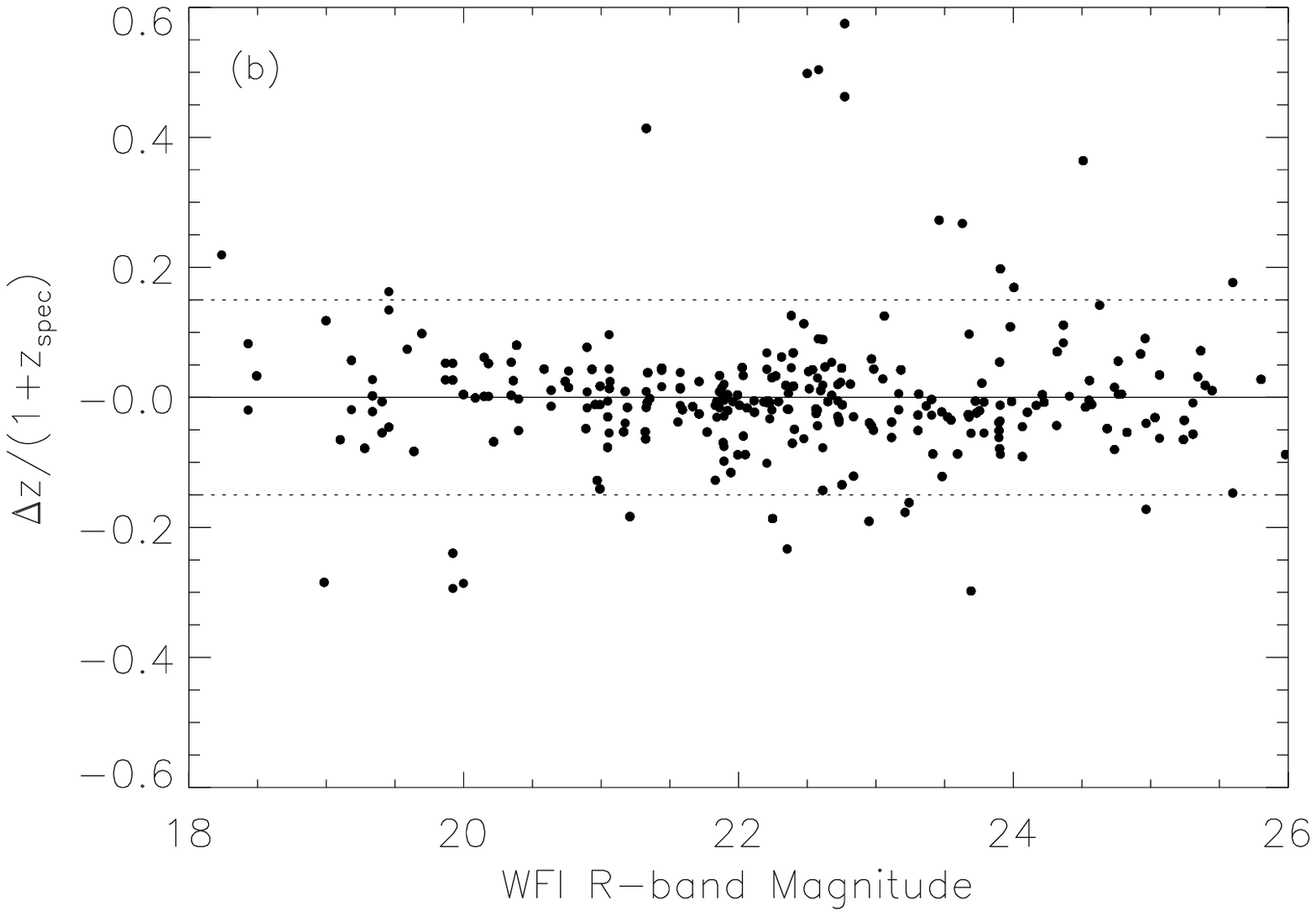}
}
\figcaption{
The photo-z accuracy, $\Delta z/(1+z_{\rm spec})$, vs. WFI $R$-band magnitude 
for ({\it a}) the spectroscopic sample and ({\it b}) the blind-test sample.
The solid lines indicate $z_{\rm photo}=z_{\rm spec}$, and
the dotted lines represent relations of
$z_{\rm photo}=z_{\rm spec}\pm0.15(1+z_{\rm spec})$.
Note that
one data point can represent several sources in the blind-test sample ({\it b}).
The photo-z accuracy declines at faint $R$-band magnitudes.
\label{rz}}
\end{figure*}

To check the quality of the photo-z's, we compared the photo-z's
to the spec-z's for the 220 sources in the spectroscopic sample (see 
Figures~\ref{zplot}a, \ref{zhist}a, and \ref{rz}a). 
The catastrophic failures (outliers)
are defined as having \hbox{$|\Delta z|/(1+z_{\rm spec})>0.15$}, where
$\Delta z=z_{\rm photo}-z_{\rm spec}$. There are 3 outliers (1.4\%) in the 
spectroscopic sample. 
The outliers typically still have acceptable SED fitting results, including
small photo-z errors, normal reduced $\chi^2$ values, and no obvious
alternative photo-z choices, and thus cannot 
be discriminated without the \hbox{spec-z} information. Specifically,
there are two relatively bright outliers in Figure~\ref{rz}a (sources 
243 and 134, with $R\approx20.0$ and $R\approx22.8$). One of them (source 243)
is caused by the bright {\it GALEX} FUV detection ($m_{\rm FUV}=21.1$), which
should have been 
suppressed due to the Lyman break for a source at 
$z_{\rm spec}=1.03$; the FUV flux may be contributed by a few surrounding
faint sources. The other one is produced by our
luminosity prior of 
$M_B\le-18$, which pushes the \hbox{photo-z} to a larger incorrect value 
($z_{\rm spec}=0.12$ and $z_{\rm photo}$=0.64).

We also use two other indicators to evaluate the \hbox{photo-z} quality, 
the normalized median absolute deviation of $\Delta z$ ($\sigma_{\rm NMAD}$) and
the average absolute scatter ($AAS$), which are defined as
\begin{equation}
\sigma_{\rm NMAD}=1.48\times {\rm median}\left(\frac{|\Delta z-
{\rm median}(\Delta z)|}{1+z_{\rm spec}}
\right)~,
\end{equation}
and
\begin{equation}
AAS={\rm mean}\left(\frac{|\Delta z|}{1+z_{\rm spec}} \right)~.
\end{equation}
$\sigma_{\rm NMAD}$ is a robust indicator of the photo-z accuracy 
after the exclusion of outliers \citep[e.g.,][]{Maronna2006,Brammer2008}, 
while $AAS$ shows the mean scatter including the contribution from outliers.
For our spectroscopic sample, $\sigma_{\rm NMAD}=0.010$ and $AAS=0.013$. 
These parameters, along with Figure~\ref{zplot}a, indicate that 
the photo-z's for the spectroscopic sample are very reliable in terms of 
both the outlier percentage and accuracy.
We show in Figure~\ref{sedexp} the SED fitting result for 
source 312 as an example; the derived photo-z for this source 
is accurate to within 1\% 
[$|\Delta z|/(1+z_{\rm spec})=0.005$].

However, the above evaluation does not represent the quality of the 
photo-z's for 
sources in the photometric sample, because the SED templates were corrected using the 
spec-z information, and we are thus biased to get optimal fitting results for 
sources with spec-z's. Therefore, we performed a ``blind test'' to 
derive a more realistic estimate of the overall \hbox{photo-z} accuracy.
We randomly picked 3/4 of the sources in the spectroscopic sample,
and only the spec-z's of these sources were used to train the 
SED templates. The \hbox{photo-z's} of the other 1/4 of the sources are 
calculated based on the improved templates, and these are the 
blind-test sources 
as there was no prior knowledge utilized about their spec-z's. The above process
was repeated 20 times to create a statistically significant sample including
1110 blind-test sources (the 
blind-test
sample). Applying the same quality indicators above to the blind-test sample, 
we get an outlier percentage of 8.6\%, $\sigma_{\rm NMAD}=0.059$, 
and $AAS=0.065$. The results are shown in Figures \ref{zplot}b,  
\ref{zhist}b, and \ref{rz}b. 
The photo-z accuracy does not appear to degrade at high redshifts ($z\approx2$--4),
and the accuracy declines at faint $R$-band magnitudes (the outlier fractions
for the $R>22.5$ and $R\le22.5$ samples are 11.4\% and 6.1\%, respectively). 
Note that among the 1110 sources in the blind-test sample, 
some have the same
spec-z, photo-z, and $R$ magnitude (same source in the spectroscopic sample,
and different training samples do not alter the best-fit \hbox{photo-z}), 
and thus cannot be distinguished in Figures~\ref{zplot}b and \ref{rz}b
(i.e., they appear as the same data point).
The effective size of the blind-test sample should be around 220, 
as we only have this number of sources in the spectroscopic sample.

\begin{figure*}
\centerline{
\includegraphics[scale=0.5]{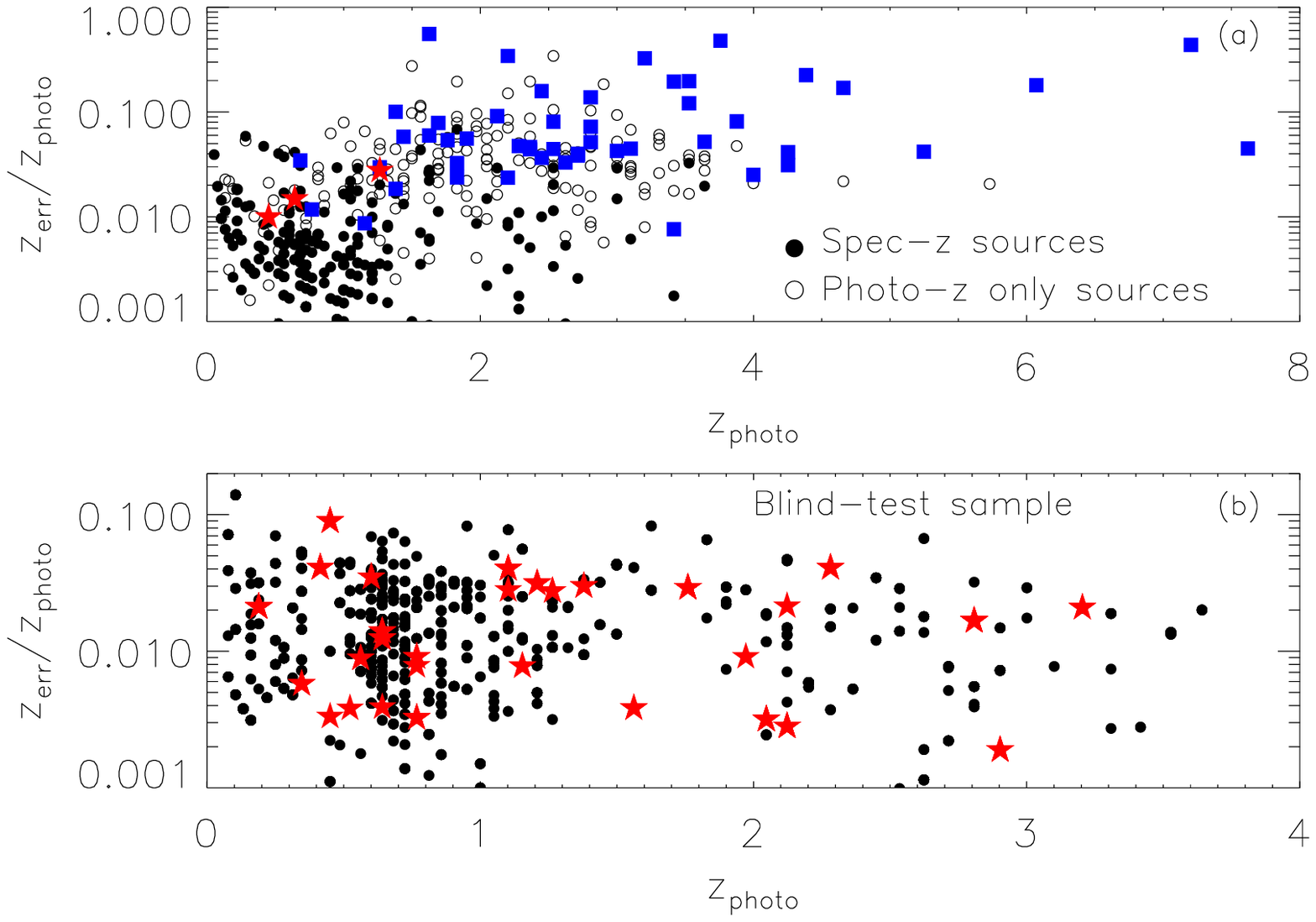}
\includegraphics[scale=0.5]{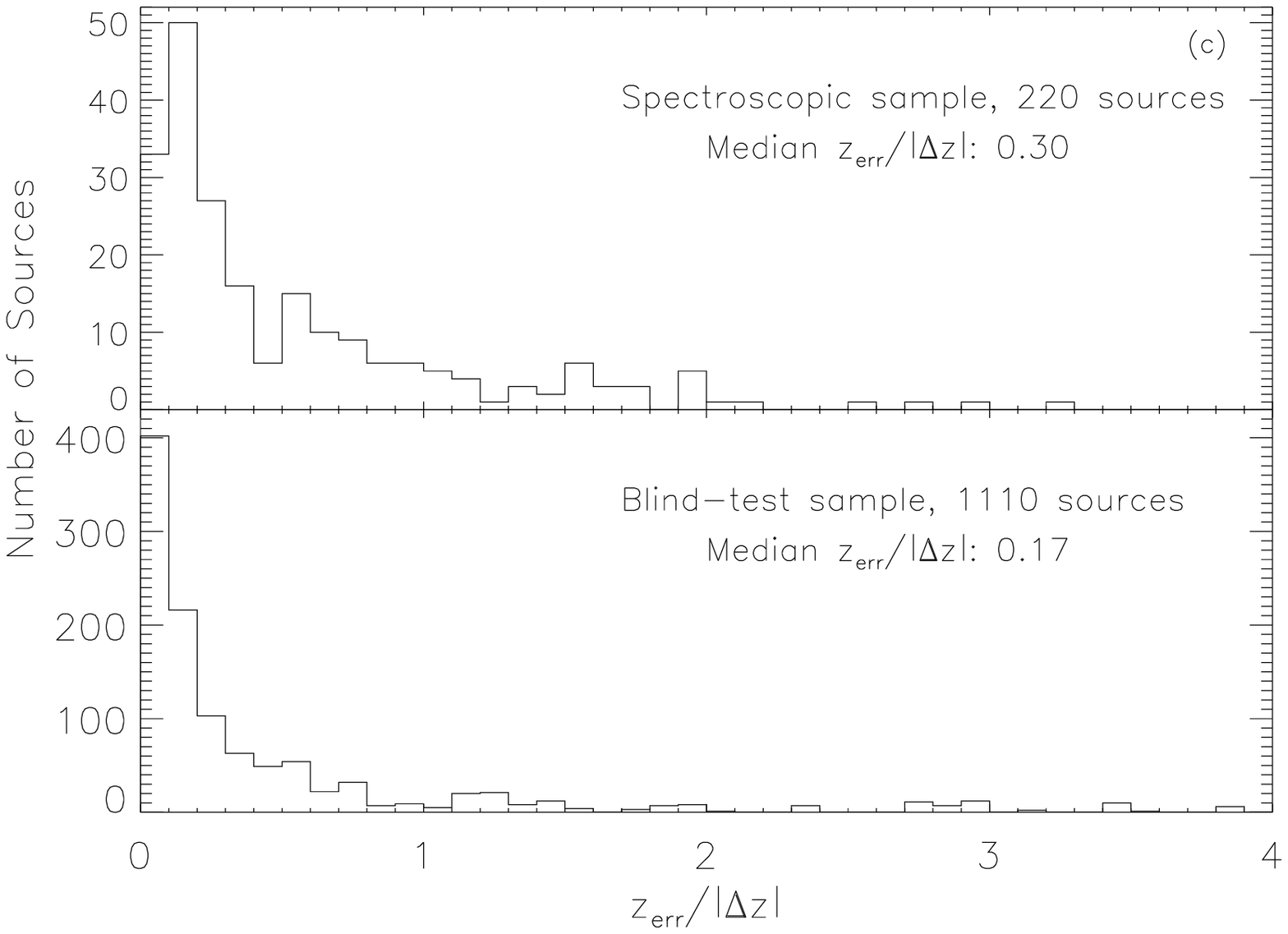}
}
\figcaption{
The distributions of the photo-z errors derived from the ZEBRA SED fitting 
for ({\it a}) the spectroscopic (filled dots) and photometric
(open dots) samples and ({\it b}) the blind-test sample. 
The ZEBRA photo-z error is 
calculated as $z_{\rm err}=(z_{\rm upper}-z_{\rm lower})/2$.
The red stars 
represent outliers in the spectroscopic and blind-test samples.
The blue squares represent sources with an alternative photo-z.
In ({\it c}), we show the distribution of the ZEBRA photo-z errors
in units of 
the real photo-z errors ($|\Delta z|$) 
for the spectroscopic sample and the \hbox{blind-test} sample.
The ZEBRA photo-z errors generally underestimate the real errors by a 
factor of $\approx3$--6. [{\it See the electronic
edition of the Journal for a color version of this figure.}] 
\label{zerr}}
\end{figure*}

In Figures~\ref{zerr}a and \ref{zerr}b, we show the distributions of the 
1~$\sigma$ photo-z errors estimated from the ZEBRA SED fitting for the 440 X-ray
sources (see Table~\ref{zcat}) and the blind-test sample, respectively.
The spectroscopic sources generally have smaller photo-z errors than the 
photometric sources, mainly because of their better multiwavelength coverage;
the ZEBRA photo-z errors also appear to decrease toward brighter 
magnitudes, for the same reason.
Sources with an alternative photo-z (blue squares in Fig.~\ref{zerr}a) 
tend to have a large photo-z error, and the outliers (red stars in
Figs.~\ref{zerr}a and \ref{zerr}b) 
cannot be distinguished
based on their photo-z errors.
The ZEBRA photo-z errors
were calculated in the two-dimensional \hbox{redshift-template} space.
However, the template plays a minor role in the error
calculation,
unless there are multiple competitive SED
fits with very close
$\chi^2$ values.
Therefore, the errors for most sources were determined
using essentially only the best-fit template, and are generally underestimates
of the real errors.
We compared the ZEBRA 1 $\sigma$ 
photo-z errors to the real errors ($|\Delta z|$)
using the spectroscopic and blind-test samples, and the results are shown in 
Figure~\ref{zerr}c. The ZEBRA photo-z errors 
only account for $\approx30\%$ and $\approx17\%$ of the 
real errors on average for the spectroscopic 
and blind-test samples, respectively.
Similar behavior has been discovered for the photo-z errors derived using 
other photo-z codes \citep[e.g.,][]{Hildebrandt2008}, in the sense that
the photo-z accuracy is not tightly correlated
with the error estimate. A more realistic estimate of the photo-z errors 
for our sources is given by the $\sigma_{\rm NMAD}$ parameter, about 1\%
for the spectroscopic sample and 6\% for the blind-test sample.

Although we give an unbiased estimate of the photo-z quality
using the blind-test sample,
we note that there are three factors that could affect the results: (1)
In the blind-test sample, only 3/4 of the spec-z sources are used to
train the templates, while in the final results, all the spec-z sources
are used in the training. Thus the photo-z quality should be somewhat
better due to the larger training sample. (2)
The quality estimate is solely
based on the spectroscopic sample, in which sources
are generally brighter and have better photometric sampling than the sources
in the photometric sample (see Fig.~\ref{filter}b).
The photo-z quality for the latter might not be as
good as what we derived here. Indeed, \citet{Hildebrandt2008} have shown
that the outlier fractions for the latter samples are typically larger by
at least a factor of two.
We performed another blind test, choosing the 182 spec-z sources with
WFI $R<24$ as the training sample, and then calculating the photo-z's of
the other 38 spec-z sources with $R\approx24$--26.
The resulting outlier fraction
is $10.5\%$ and $\sigma_{\rm NMAD}=0.067$,\footnote{The photo-z quality does not appear to 
degrade from $R>22.5$
to $R>24$ (see Figure~\ref{rz}b), maybe due to the 
small sample size of
$R>24$ sources. Therefore the outlier fraction here is 
comparable to that (11.4\%) for the $R>22.5$ sources in the blind-test sample
above.} 
which are not greatly increased relative to the 8.6\% outlier fraction and 
$\sigma_{\rm NMAD}=0.059$
for the blind-test sample above. 
Note that these 38 faint spec-z sources are still
significantly brighter than the sources 
in the photometric sample on average (see, e.g.,
Figure~\ref{filter}b); therefore an outlier fraction of $\approx15$--25\%
is probable for the faint sources ($R\ga26$) in the photometric sample.
(3) The spec-z's were used when we selected the
SED templates. This should not affect the blind test significantly,
as we chose a large and relatively complete SED library.

In summary, among the 220 photo-z's in the photometric sample, 
we expect there are $\approx9\%$ outliers for the $\approx110$
relatively brighter sources ($R\la26$), and the outlier fraction will rise
to $\approx15$--25\% for the fainter sources ($R\ga26$). 
The typical photo-z
accuracy is \hbox{$\approx6$--7\%} ($\sigma_{\rm NMAD}$).
The outlier fraction and photo-z
accuracy do not appear to have a redshift dependence (for $z\approx0$--4).
Among the faint $R\ga26$ sources, 20 objects do not have any 
optical counterparts (see Table~\ref{zcat}, 
including the five $z_{\rm photo}>5$ sources shown in 
Fig.~\ref{f09}a), and the photo-z's were calculated using only the 
IR/NIR photometric data and the optical
upper-limit information. Caution should be taken when using these 
photo-z's. Moreover, the 
alternative photo-z's provided for the 49 sources may occasionally 
be a better choice than the best-fit photo-z's (e.g., for some of the 
$z_{\rm photo}>4$ sources). 

Recently, 
the GOODS/VIMOS spectroscopic campaign has provided
two additional secure spec-z's for our X-ray sources \citep{Balestra2009}, 
$z_{\rm spec}=0.740$ for source 175 and 
$z_{\rm spec}=2.586$ for source 372.
Their photo-z's in our catalog are 0.56 and 2.71, with accuracies of
$|\Delta z|/(1+z_{\rm spec})\approx10\%$ and 3.6\%, respectively.

\subsection{Comparison with Other AGN Photometric-Redshift Accuracies}

The photo-z quality of the spectroscopic sample (1.4\% outlier and 
$\sigma_{\rm NMAD}=0.01$) is comparable to those
of recent \hbox{photo-z} studies
for the COSMOS optical \citep[e.g.,][]{Ilbert2009} and
X-ray \citep{Salvato2009} sources.
Compared to the blind-test result (46 sources)
for the COSMOS \hbox{X-ray} sources \citep{Salvato2009} with 10.9\%
outliers and $\sigma_{\rm NMAD}\approx0.023$, our result has a slighter
smaller fraction (8.6\%) of outliers but a larger $\sigma_{\rm NMAD}$ (0.059).
Note that the CDF-S and COSMOS X-ray sources are two
different populations. The $\approx2$~Ms CDF-S sources have an average 
\chandra\ effective exposure of $\approx1$~Ms, while the COSMOS sources have
an average effective {\it XMM-Newton} exposure of $\approx50$~ks. In our
spectroscopic sample, $\approx38\%$ of the sources have {\it HST}
$i>22.5$, and in the COSMOS spectroscopic sample, only $\approx10\%$ of
the sources have Canada--France--Hawaii Telescope 
$i^*>22.5$ \citep{Salvato2009}. 
Our \hbox{photo-z} quality is at least
comparable to that for the COSMOS X-ray sources, considering that we are
dealing with a fainter and thus more challenging sample.
One advantage in the photo-z work for the COSMOS X-ray sources is
that a correction was made for optical AGN variability, 
which helps to reduce the outlier fraction and $\sigma_{\rm NMAD}$
\citep{Salvato2009}. Given the nature of the photometric data in the CDF-S,
i.e., multiwavelength observations are not separated in well-defined epochs,
source variability
is difficult to take into account. However, we also note that for 
the moderate-luminosity AGNs in our sample, the SEDs are often dominated 
by host-galaxy light (see Fig.~\ref{f09}b), and thus AGN variability
may play a smaller role for our sources than for the brighter COSMOS AGNs.

Compared to the photo-z quality for the 342 X-ray sources in the $\approx1$ Ms CDF-S
\citep{Zheng2004}, with $\approx15\%$ outliers and 
$\sigma_{\rm NMAD}\approx0.08$--0.1,\footnote{We also checked the 
photo-z quality of the \citet{Zheng2004} sources using 31 spec-z sources that
did not have spec-z information in \citet{Zheng2004}. The resulting outlier
fraction is $\approx26\%$, and the outlier-excluded accuracy is $\sigma_{\rm NMAD}\approx0.08$.}
or the photo-z quality of the 429
\hbox{X-ray} sources in the $\approx2$ Ms CDF-N \citep{Barger2003}, with 94\% of the photo-z's
accurate to within $\approx25\%$,
our photo-z accuracy estimated based upon the blind-test sample is significantly
($\ga50\%$) improved.

\subsection{AGN Classification and Best-Fit SED Templates}

The intrinsic hydrogen column density $N_{\rm H}$ estimated in \S\ref{zresults} 
(assuming an underlying $\Gamma=1.8$) can provide
a basic AGN classification from the X-ray point of view, i.e., ``absorbed'' with
$N_{\rm H}\ge10^{22}$~cm$^{-2}$ or
``unabsorbed'' with $N_{\rm H}<10^{22}$~cm$^{-2}$.
This method has limited diagnostic utility because of the complexity of
AGN X-ray spectra. However, it suggests that $\approx256$ ($\approx72\%$)
of the 354 AGNs in the spectroscopic and photometric samples are absorbed, 
which is in general agreement with the
$\approx60$--75\% absorbed AGN fraction in the $\approx1$ Ms CDF-S
and $\approx2$ Ms CDF-N
based on X-ray spectral analysis \citep[e.g.,][]{Bauer2004,Tozzi2006}.

We investigated the best-fit SED templates selected
for the 440 sources; about 98\%
of them are improved templates from the template-training procedure,
indicating the importance of the training step.
Of the 265 input templates (including their trained varieties), 105 are
selected as the best-fit SED template for multiple sources
and 107 are never used; the most frequently used template (a
starburst-galaxy
template) has been chosen for 22 sources.
The best-fit SED templates can also provide
a basic SED-based classification of the X-ray sources.
Of the 440 source SEDs, 212 are best fit by galaxy templates,
and the others
are fit by AGN or hybrid templates, including 84 type
1 AGNs and 144 type 2 AGNs.
For sources that are detected in only a few or none of the optical bands,
this classification may be highly uncertain.

AGNs identified in the X-ray do not completely agree with the SED fitting
results, as expected based upon disagreements between X-ray based and
optical-spectroscopy based classification schemes \citep[e.g.,][]{Matt2002,Trouille2009}.
For example, of the 303 AGNs
with detections in at least
ten \hbox{UV--to--IR} bands,
125 (41.3\%) are
best fit by galaxy templates. Despite the uncertainties in the SED fitting
and the calculation
of X-ray luminosities, some of these sources are likely
the X-ray bright optically
normal galaxies (XBONGs; e.g., \citealt{Comastri2002}), which
have high X-ray luminosities but do not appear as AGNs in the optical. The
nature of XBONGs remains somewhat mysterious: the properties of
XBONGs are usually explained by heavy obscuration covering
a large solid angle \citep[e.g.,][]{Matt2002,Rigby2006,Civano2007},
dilution by host galaxies
(e.g., \citealt{Moran2002}; \citealt{Severgnini2003};
\citealt{Trump2009}; see also Fig.~\ref{f09}b),
or a
radiatively inefficient accretion flow
\citep[RIAF; e.g.,][]{Yuan2004,Trump2009}.

\begin{figure*}
\centerline{
\includegraphics[scale=0.5]{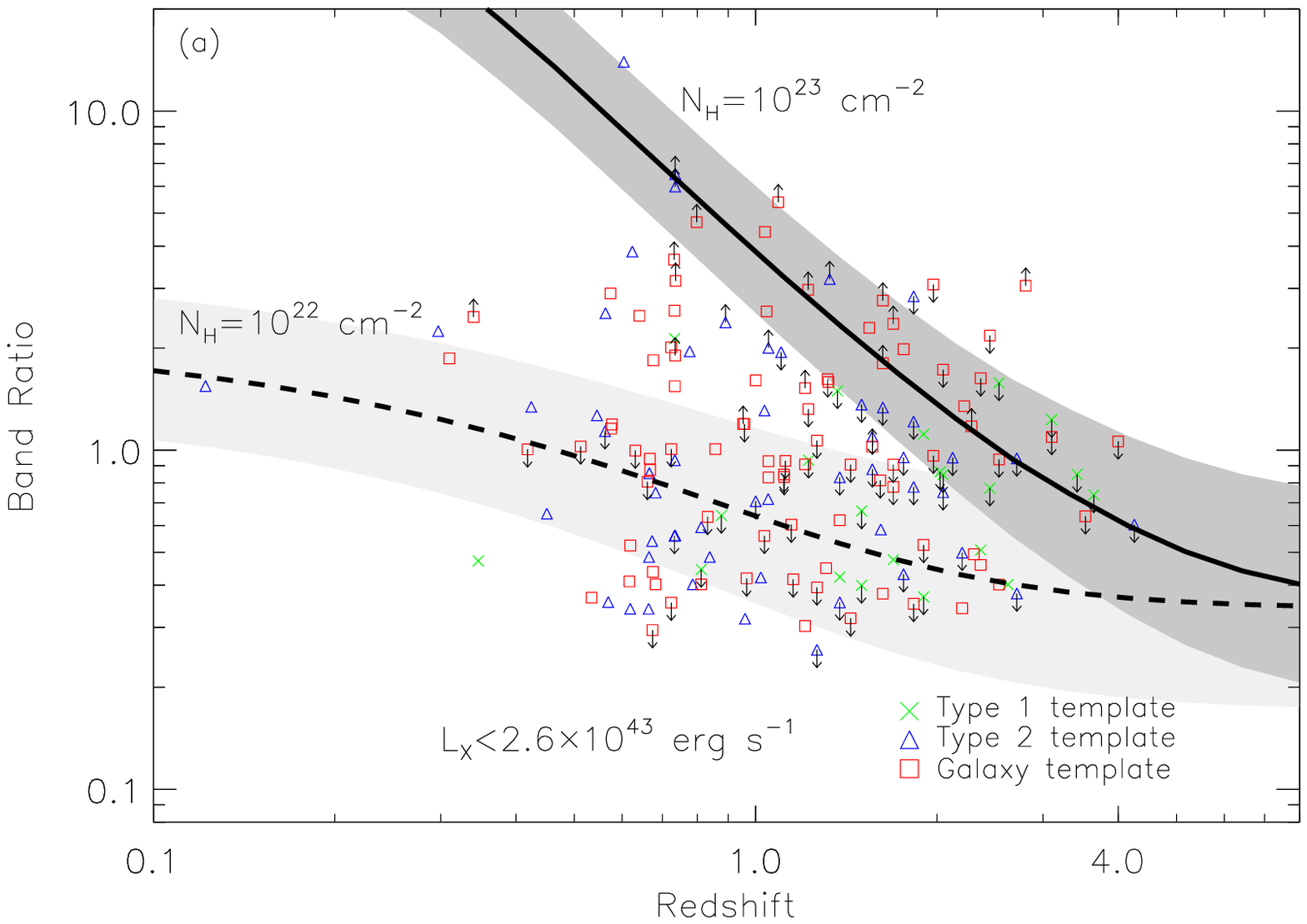}
\includegraphics[scale=0.5]{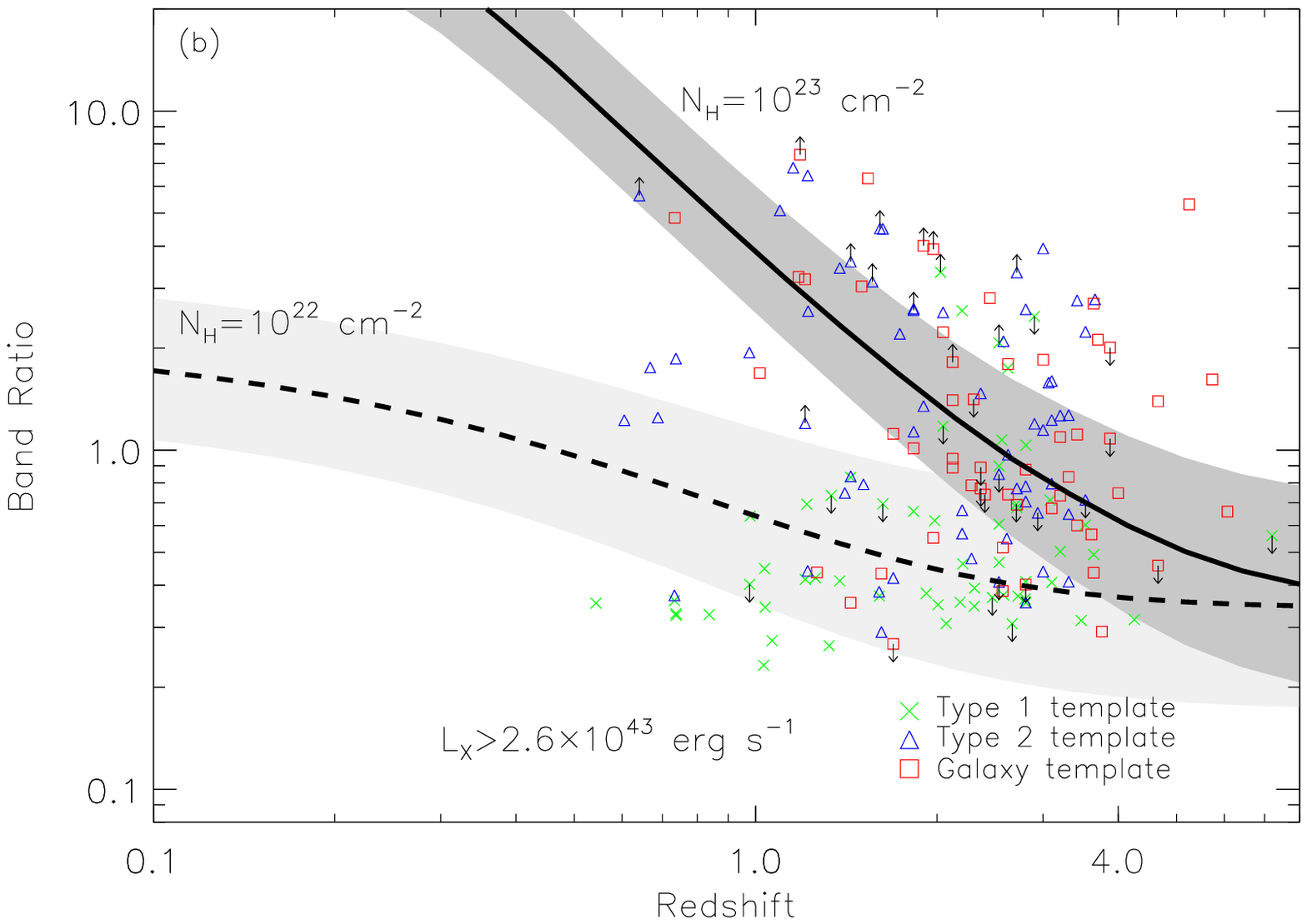}
}
\figcaption{
X-ray band ratio vs. redshift for the ({\it a}) fainter and ({\it b})
brighter halves of AGNs, separated by the 
median X-ray luminosity,
$L_{\rm 0.5\textrm{--}8.0~keV,rest}=2.6\times10^{43}$~\lum.
Band ratios are defined as the ratio of count rates
between the hard and soft bands.
The green crosses, blue triangles, and red squares are
sources whose SEDs are best-fit by type 1, type 2, and galaxy
templates, respectively.
Upper limits (lower
limits) on the band ratios are shown when the sources are not
detected in the hard band (soft band). The errors on the band ratios
are not shown for clarity; the median error is $\approx0.16$.
The solid (dashed) curve and the dark (light) shaded region show
the expected band ratios
for AGNs with power-law photon indices $\Gamma=1.8\pm0.5$ and
intrinsic absorption $N_{\rm H}=10^{23}$~cm$^{-2}$
($N_{\rm H}=10^{22}$~cm$^{-2}$).
The \hbox{X-ray} (unabsorbed versus absorbed)
and SED-based (type 1 versus non-type 1) classification schemes
agree better for luminous AGNs; $\approx 86\%$ of the non-type 1
sources are absorbed, and about half of the type 1 sources are unabsorbed.
[{\it See the electronic
edition of the Journal for a color version of this figure.}]
\label{hratio}}
\end{figure*}

An apparent inconsistency also exists between
the X-ray (unabsorbed versus absorbed)
and SED-based (type 1 versus non-type 1, including type 2 and galaxies)
classification schemes.
Figure~\ref{hratio} displays the \hbox{X-ray} band ratio
versus redshift for the 354 AGNs.
The SED-based classification is color
coded, while the X-ray classification is indicated by the theoretical
band-ratio--redshift track.
We study the classifications in two X-ray luminosity bins, divided by
our median X-ray luminosity,
$L_{\rm 0.5\textrm{--}8.0~keV,rest}\approx2.6\times10^{43}$~\lum.
We consider only the 303 AGNs with detections in at least
ten \hbox{UV--to--IR} bands. In the low-luminosity (high-luminosity) bin,
42\% (55\%) of the type 1 AGNs are unabsorbed,
64\% (86\%) of the type 2 AGNs are absorbed,
and 71\% (87\%) of the AGNs fit by galaxy templates are absorbed.
For relatively luminous \hbox{X-ray} sources, the two classification schemes
tend to agree better, probably due to less fractional
contamination from host-galaxy light.\footnote{
\citet{Salvato2009} also made a similar comparison of the SED
versus X-ray classifications for the COSMOS X-ray sources, and found 
a generally better agreement ($\approx80\%$), probably
due to the fact that COSMOS samples brighter X-ray fluxes than
the CDF-S.} 
It appears that most of the non-type 1 sources ($\approx 86\%$ for
luminous X-ray sources) are absorbed,
while only about half of the type 1 sources are unabsorbed.
The comparison between the X-ray and optical-spectroscopy classification
schemes in \citet{Trouille2009} shows that $\approx80\%$ of the
broad-line AGNs are unabsorbed and $\approx67\%$ of the non-broad-line AGNs
are absorbed. Therefore, the \hbox{X-ray} versus optical-spectroscopy
classification schemes agree better for unabsorbed/type 1 AGNs, compared to the
X-ray versus SED classification schemes.
Note that the uncertainties in the simple
X-ray classification scheme above are plausibly responsible for some
of the classification discrepancy;
detailed X-ray spectral analysis will
provide a better X-ray classification and give more insight into the problem
of X-ray versus SED classification inconsistency.

\section{DISCUSSION}

\subsection{Nature of the Unidentified X-ray Sources}\label{noiddis}
There are 16 unidentified sources, 
eight of which are probably related to a star (one source), 
galaxy groups and clusters (four sources), off-nuclear 
sources (two sources), or any of a number of galaxies at the edge 
of the CDF-S field (one source; see Col. 13 of Table \ref{cat}).
For the other eight X-ray 
sources (flagged with ``NoID'' in Col. 13 of Table \ref{cat}; 
see Figure~\ref{coverage} for their locations),
six were detected only at a false-positive 
threshold of $1\times10^{-6}$, and two (sources 49 and 345) at $1\times10^{-8}$.
At a threshold of $1\times10^{-8}$, we expect less than 0.2 false detections
(see \S3.2 of L08).
We visually checked the X-ray images of sources 49 and 345 and confirmed
their apparent validity. We also investigated the \chandra\ events for these two sources 
and excluded the possibility of contamination from short-lived
cosmic-ray afterglows. 
Therefore sources 49 and 345 are likely genuine X-ray sources. 
Source 49 is outside of the GOODS-S region. It was detected in all
three X-ray bands ($\approx117$ \hbox{full-band} counts) 
and was also present in the $\approx1$ Ms CDF-S 
\citep{Giacconi2002,Alexander2003}. 
It is a hard X-ray source with a band ratio of 1.5 (corresponding to 
an intrinsic column density of 
$N_{\rm H}\approx6\times10^{23}$~cm$^{-2}$ assuming $z=5$ and 
$\Gamma=1.8$). Its 0.7--6 keV X-ray spectrum shows no 
obvious emission-line features.
The X-ray--to--optical ($z$-band) flux
ratio is $F_{\rm 0.5\textrm{--}8.0~keV}/F_z\ga50$. 
There appears to be marginal 3.6, 4.5, and 8.0~$\mu$m emission
from the X-ray
position of source 49.
However, there is a relatively bright [$m(3.6~\mu{\rm m})=21.2$] IR source
nearby (2.3\arcsec\ away); even if the emission is from source 49, it
is completely blended with the bright source.
There is also a faint VIMOS $U$-band source ($U=28.70\pm0.22$) 
1.3\arcsec\ away from source 49,
which is not detected in any of the other bands. It was not considered a 
counterpart by the likelihood-ratio matching.
Source 345 is inside the GOODS-S region and is only detected in the 
soft band with $\approx6$ counts. The detection is likely significant 
because there are five counts landing on a single pixel. 
Its \hbox{X-ray--to--optical} flux
ratio is $F_{\rm 0.5\textrm{--}8.0~keV}/F_z\ga10$. 
The nearest ONIR source to source 345 is $\approx2.0\arcsec$ 
away.

The properties of sources 49 and 345 are comparable to the 
EXOs in
\citet{Koekemoer2004}, except that they do not have any clear detections in the 
IR or NIR bands (which EXOs generally have) and source 345 is relatively X-ray weak.
Most of the EXOs are suggested to be dusty galaxies
at moderate redshifts ($z\approx2$--3), and a small fraction could be
AGNs at very high redshifts ($z>5$; e.g., \citealt{Koekemoer2004,Mainieri2005}).
Given the apparent 
non-detection of these two sources in the deep IR and NIR images, 
they 
are unlikely to be dusty galaxies
at moderate redshifts (i.e., their host galaxies should be detected). 
Therefore, sources 49 and 345 are more likely to be
AGNs at very high redshifts that appear blank in all the
current optical--to--IR catalogs due to finite-sensitivity effects.
Using the luminosities and spec-z's or photo-z's of our X-ray sources, and the 
depths of the \hbox{optical--to--IR} catalogs,
we estimate that $\approx45$ ($\approx10\%$) of the sources would appear blank in
all these bands if 
placed at $z=5$, and $\approx10$ of them would still be detected in the X-ray
if at $z=5$.
There is also the possibility that source 345 is an  
isolated neutron star, such as the so-called ``Dim Isolated Neutron Stars'' 
\citep[e.g.,][]{Kaplan2008}, which emit thermal radiation with temperatures of 
$\approx0.1$~keV.
This is not likely the case for source 49 
because of its hard X-ray emission. 
Isolated neutron stars with faint X-ray fluxes such as source 345
would appear blank at optical--to--IR wavelengths, because of
the high ($\approx10^3$--$10^4$) X-ray--to--optical flux ratios of these sources.
Given the X-ray sensitivity of the CDF-S (L08) and the number counts of
isolated neutron stars \citep[e.g.,][]{Popov2000,Popov2010},
we expect $\approx0.2$ detectable isolated neutron stars in the CDF-S field.
Very deep NIR imaging or
spectroscopic observations will help elucidate the nature of sources 49 and 345.

The six sources (XIDs 93, 106, 210, 280, 344, and 353) detected at a
\hbox{low-significance} level of $1\times10^{-6}$ and without identifications could be spurious X-ray detections.
Visual inspection indicates that most of these sources appear only marginally 
above the background.
At a detection threshold of $1\times10^{-6}$, we conservatively estimated 
that there are $\approx18$ false detections (see \S3.2 of L08), 
and the true number is likely
$\approx2$--3 times less (\citealt{Alexander2003}). 
Spurious \hbox{X-ray} detections could also exist in the 
identified source sample because of chance correlations. 
We estimate that the average possibility of matching a 
randomly placed $1\times10^{-6}$ X-ray 
source to an ONIR source is $\approx11\%$, using the same bootstrapping
technique described in \S\ref{results}.\footnote{Essentially, this is 
an estimate of the false-match probability for 
X-ray sources detected at only the
$1\times10^{-6}$ significance level. The 
false-match probability is higher
than that for the entire X-ray sample ($\approx6.2\%$), mainly because the
positional uncertainties of these X-ray sources are relatively larger.}
If the six unidentified X-ray sources are indeed spurious, we expect less than
one [$6\times11\%/(1-11\%)$] spurious X-ray source in our identified
source sample. 
Therefore, there are up to $\approx7$ spurious X-ray detections, six 
unidentified and one falsely identified. This number is roughly consistent
with the expected value ($\approx2$--3 times less than 18).
If any of these six low-significance and unidentified sources 
is a real X-ray detection, it could have a similar nature as sources 49 and 345.

\subsection{Future Prospects for Improved Source Identification}

Besides the unidentified X-ray sources, uncertainties are still present for the
442 identified sources (excluding the four additional likely cases), as there is a 
$\approx6.2\%$ false-match probability and there are 21 primary counterparts
from the MUSYC or SIMPLE catalogs with relatively less-accurate 
(0.2--0.3\arcsec)
positions.
In order to characterize the nature of all the X-ray sources in the 
$\approx2$ Ms CDF-S, efforts should focus on 
obtaining ({\it a}) better positions for the counterparts and ({\it b})
a lower overall false-match probability.
Better counterpart positions generally require deeper optical 
or NIR observations.
Comparing the current deepest optical (GOODS-S $z$-band) and NIR (MUSIC $K$-band)
catalogs, we consider the NIR band to be probably the most 
beneficial to push deeper, because of its
higher counterpart recovery rate, 
X--O fraction, and lower false-match probability 
(see Table~\ref{matchsum}) at the current depth. 
Deep {\it HST} WFC3/IR surveys or the forthcoming 
{\it James Webb Space Telescope} ({\it JWST\/}) will provide 
unprecedented NIR/IR data for source identification.
Note that a deeper optical or NIR catalog will generally result in a larger
false-match probability.

There are at least two possible approaches toward a smaller overall false-match 
probability: deeper \hbox{X-ray} observations and deeper radio observations.
The probability distribution of angular separations, $f(r)$, is largely
controlled by the \hbox{X-ray} positional errors (see Eq. 2) except when matching
to the IR sources. Additional 
X-ray observations will improve the \hbox{X-ray} positional accuracy and thus
reduce the false-match probability. For example, as the $\approx2$ Ms CDF-S is 
scheduled to be
expanded to $\approx4$ Ms, the factor of $\approx2$ increase in counts will 
reduce the areas of source
positional error regions by $\approx30\%$ on average, 
reducing the number of spurious matches (at the current flux limit) 
by about this same factor.
The VLA radio catalog has the highest \hbox{X--O} fraction ($\approx28\%$) and lowest 
false-match probability (see Table~\ref{matchsum}). However, it also has 
the lowest counterpart recovery rate ($\approx20\%$). 
To examine if the high X--O fraction
of the radio catalog is intrinsic or rather just related to the 
low recovery rate (i.e., limited
catalog depth), we compared it to the X--O fractions of optical and NIR
catalogs at similar effective depth.
We limit the GOODS-S $z$-band catalog and MUSIC $K$-band 
catalog to brighter magnitudes ($z<21.0$ and $K<19.8$) until the recovery 
rates of these two 
catalogs are the same as that of the radio catalog ($\approx20\%$).
The resulting X-O fractions are only $\approx13\%$ and $\approx17\%$ 
for these two catalogs, respectively. This suggests the X-ray sources 
are more strongly associated with the 
radio sources than optical or NIR sources at
the current depth of the radio catalog, and thus deeper radio observations
might well provide more reliable matches than the GOODS-S or MUSIC catalogs.
The current radio catalog has a 5~$\sigma$ limiting flux of $\approx40~\mu$Jy.
Given the VLA 1.4 GHz number counts estimated in \citet{Owen2008}, there will
be $\approx20$ times more radio sources at a flux limit of $5~\mu$Jy.
This radio source density (about 30 sources arcmin$^{-2}$) 
is still $\approx2$--4 times smaller than those of the optical and NIR
catalogs, and we expect that such deeper radio catalogs can
provide more reliable matches and thus reduce the overall
false-match probability. The flux limit of $5~\mu$Jy can be easily 
achieved by facilities such as the Expanded Very Large Array (EVLA).
The Atacama Large Millimeter Array
(ALMA) at shorter wavelengths may also help with the goal of obtaining
more reliable counterparts in a similar way.

\section{SUMMARY OF RESULTS}
\begin{enumerate}
\item
We presented multiwavelength identifications of the 462 X-ray sources in 
the main \chandra\ source catalog for the $\approx2$~Ms CDF-S. 
The optical--to--radio catalogs used for source matching include
the WFI $R$-band, GOODS-S $z$-band, GEMS $z$-band, \hbox{GOODS-S} MUSIC $K$-band,
MUSYC $K$-band, SIMPLE 3.6 $\mu$m, and VLA 1.4 GHz catalogs. 
The matching results are combined to create lists of primary
and secondary counterparts.
We identified
reliable counterparts for 442 (95.7\%) of the X-ray sources, with an expected 
false-match probability of $\approx6.2\%$; these are the best
identifications for sources at these X-ray fluxes. 
We also associated four additional likely counterparts given their small
positional offsets to the relevant X-ray sources.

\item
The likelihood-ratio method was used for source matching,
which significantly reduces the \hbox{false-match} probability at faint
magnitudes compared to the \hbox{error-circle} matching method. For example,
the \hbox{false-match} probabilities at $R=26$ are $\approx20\%$ and $\approx28\%$
for the likelihood-ratio and error-circle methods, respectively.
The resulting matches may differ by $\approx20\%$ for sources with $R>25.5$,
where we expect to find some of the most interesting objects in the deep X-ray
surveys. The likelihood-ratio matching technique can also generate 
multiple counterpart candidates and calculate their relative reliability 
parameters, which are useful when considering 
multiwavelength identifications.

\item
Among the optical--to--radio multiwavelength catalogs used for identification,
the \hbox{false-match} probabilities range from $\approx1$--9\%, and the
counterpart recovery rates range from \hbox{$\approx20$--80\%}. Given the 
current depths of the catalogs, the NIR catalogs (e.g., the MUSIC $K$-band
catalog)
can provide a large fraction of the counterparts for the X-ray sources with
a relatively low false-match probability. Meanwhile, the 
radio catalog can provide the most secure counterparts with almost no 
spurious matches. Even after limiting the optical/NIR catalogs to similar
effective depth as the radio catalog, the radio sources still have a significantly
stronger association with the X-ray sources (measured by the X--O fraction 
parameter). Therefore, obtaining deeper radio observations will 
considerably benefit the matching reliability. Additional X-ray observations
will reduce the X-ray positional errors and thus also improve the 
identification reliability.

\item
Of the 16 unidentified X-ray sources, 
eight are probably related to a star (one source), 
galaxy groups and clusters (four sources), off-nuclear 
sources (two sources), or any of a number of galaxies at the edge 
of the \hbox{CDF-S} field (one source). Another six sources detected at 
a {\sc wavdetect} 
false-positive threshold of $1\times10^{-6}$ may be spurious 
X-ray detections. The remaining two objects are likely genuine X-ray sources.
Their X-ray--to--optical properties are comparable to the EXOs, 
though without any IR or NIR detections. 
They are likely to be AGNs at very high redshifts ($z>5$), 
or one of them (source 345)
could be an isolated neutron star.

\item
We constructed a photometric catalog for the 446 identified X-ray
sources
including up to 42 bands of \hbox{UV--to--IR} data.
We performed source deblending carefully for
$\approx10\%$ of the sources in the IR bands and a few percent in the
optical and NIR bands, obtaining the best-ever
photometric data for these sources. \hbox{High-quality} photo-z's
were then derived using the ZEBRA SED fitting code and a set of 265
galaxy, AGN, and galaxy/AGN hybrid templates.
The ZEBRA training procedure  
corrects the SED templates to
best represent the SEDs of real sources at different redshifts,
which reduces template mismatches
and improves the photo-z accuracy.
Catalogs of photometric data
and photo-z's are provided.
We collected secure spec-z's for 220 of the X-ray sources. 
It appears that the majority of the X-ray sources without spec-z's
are AGNs at relatively high redshifts ($z\approx1$--4).

\item
The photo-z's are extremely accurate for sources with spec-z's.
There are only 3 outliers out of 220 sources, and the photo-z's
are accurate to within $\approx1\%$ ($\sigma_{\rm NMAD}$).
We performed blind tests to
derive a more realistic estimate of the overall photo-z accuracy for
sources without spec-z's. 
We expect there are $\approx9\%$ outliers for the $\approx110$
relatively brighter sources ($R\la26$), 
and the outlier fraction will increase
to $\approx15$--25\% for the $\approx110$ fainter sources ($R\ga26$). 
The typical photo-z
accuracy is $\approx6$--7\% ($\sigma_{\rm NMAD}$).
The outlier fraction and \hbox{photo-z}
accuracy do not appear to have a redshift dependence (for \hbox{$z\approx0$--4}).
The photo-z quality is comparable to that of some recent photo-z's derived
for AGNs at brighter X-ray fluxes, and is significantly ($\ga50\%$)
improved compared to that of the photo-z's for the X-ray sources in the 
$\approx1$ Ms CDF-S or $\approx2$ Ms CDF-N.
\end{enumerate}

~\\
We acknowledge financial
support from CXC grant SP8-9003A/B (BL, WNB, YQX, MB, FEB, DR), 
NASA LTSA grant
NAG5-13035 (WNB), the Royal Society and the Philip Leverhulme Fellowship
Prize (DMA),
and ASI-INAF grants I/023/05/00 and I/088/06 (AC).
We thank R.~Feldmann, R.~Gilli, H.~Hao, R.C.~Hickox, 
K.I.~Kellermann, N.A.~Miller, and G.~Pavlov for helpful discussions.
We are grateful to A.~Finoguenov, A.~Grazian, N.A.~Miller, and M.~Nonino 
for kindly providing their catalogs.

\clearpage

\begin{turnpage}
\begin{deluxetable}{lccccccccccccccccc}

\tabletypesize{\footnotesize}
\tablewidth{0pt}
\tablecaption{Summary of the Likelihood-Ratio Matching Parameters and Results}

\tablehead{
\colhead{Catalog} &
\colhead{Det. Thresh} &
\colhead{Depth} &
\colhead{Solid Angle} &
\colhead{$N_{\rm o}$} &
\colhead{$\sigma_{\rm o}$} &
\colhead{$L_{\rm th}$} &
\colhead{$R$} &
\colhead{$C$} &
\colhead{$N_{\rm X}$} &
\colhead{$N_{\rm ID}$} &
\colhead{$N_{\rm NoID}$} &
\colhead{$N_{\rm Multi}$} &
\colhead{$N_{\rm False}$} &
\colhead{False \%} &
\colhead{Recovery \%} &
\colhead{X--O \%} &
\colhead{$N_{\rm Pri}$} \\
\colhead{(1)}         &
\colhead{(2)}         &
\colhead{(3)}         & 
\colhead{(4)}         &
\colhead{(5)}         &
\colhead{(6)}         &
\colhead{(7)}         &
\colhead{(8)}         &
\colhead{(9)}         &
\colhead{(10)}        &
\colhead{(11)}        &
\colhead{(12)}        &
\colhead{(13)}        &
\colhead{(14)}        &
\colhead{(15)}        &
\colhead{(16)}         &
\colhead{(17)}         &
\colhead{(18)}        
 }

\startdata
WFI $R$ & 2 $\sigma$ &27.3 &1420&30\,345& 0.1\arcsec &0.55 & 0.97 & 0.73 & 462 & 344 & 118 & 2 & 32 & 9\% &67.5\%& 1.0\%&19 \\
GOODS-S $z$ & 0.6 $\sigma$&28.2&160& 33\,955&0.1\arcsec &0.80 & 0.93 & 0.82 & 311 & 259 & 52 & 8 & 23& 9\% &75.9\%& 0.7\%&220\\
GEMS $z$ &  1.7 $\sigma$ &27.3&830&22\,016&0.1\arcsec &2.35 & 0.97 & 0.67 & 462 & 312 & 150 & 5 &14 & 4\% &64.5\%& 1.4\%&89 \\
MUSIC $K$ & 1 $\sigma$ &23.8&140& 13\,595&0.1\arcsec & 1.30 & 0.93 & 0.84 & 262 & 223 & 39 & 4 &8 & 4\% &82.1\%&1.6\%&14\\
MUSYC $K$ & 23.5/arcsec$^2$&22.4&970& 6998&0.2\arcsec & 0.85 & 0.99 & 0.70 & 462 & 326 & 136 & 0 & 12& 4\% &68.0\%&4.5\%&9 \\
SIMPLE 3.6 $\mu$m & 2 $\sigma$ &23.8&1640&22\,095&0.3\arcsec & 0.20 & 0.99 & 0.89 & 462 & 414& 48& 0 & 32 &8\%&82.7\% & 1.7\%&12\\
VLA 1.4 GHz & 5 $\sigma$& 19.9  &1170&338&0.1\arcsec & 2.25 &0.99 &0.20 &462 &94 &368 & 0 & 0.5 & 1\% &20.2\%& 27.7\%&83 \\
\enddata

\footnotesize

\tablecomments{Col. (1): ONIR catalog.
References.--- WFI and GOODS-S: \citet{Giavalisco2004}; GEMS: \citet{Caldwell2008};
MUSIC: \citet{Grazian2006}; MUSYC: \citet{Taylor2009}; SIMPLE: \citet{Damen2010};
VLA: \citet{Miller2008}.
Col. (2): The minimum threshold used for source detection in the
catalog. Note that in some cases multiple searches have been performed with
different thresholds for deblending purposes.
Col. (3): Catalog depth in AB magnitude. 
The AB magnitudes
for radio sources were converted from the radio flux densities,
\hbox{$m({\rm AB})=-2.5\log(f_\nu)-48.60$}.
Col. (4): Catalog solid-angle coverage in units of arcmin$^2$. 
The GEMS catalog includes GOODS-S v1 data. All the catalogs cover
the entire $\approx2$ Ms CDF-S ($\approx436$ arcmin$^{2}$) except for the GOODS-S and 
MUSIC surveys.
Col. (5): Number of ONIR sources in the $\approx2$ Ms CDF-S region.
Col. (6): 1 $\sigma$ positional error of the ONIR sources.
Col. (7): Threshold value for the likelihood ratio to discriminate between
spurious and real identifications. The threshold value is catalog dependent,
and generally scales
with the typical values of likelihood ratios (see Eq. 1), which usually
increase when the catalog depth or positional errors decrease.
Col. (8): Sample reliability. See \S\ref{method} for details.
Col. (9): Sample completeness. See \S\ref{method} for details.
Col. (10): Total number of \hbox{X-ray} sources that are within the 
coverage of the ONIR catalog.
Col. (11): Number of \hbox{X-ray} sources identified with at least one 
ONIR counterpart in this catalog.
Col. (12): Number of \hbox{X-ray} sources not identified, which equals
Col. (10) minus Col. (11).
Col. (13): Number of \hbox{X-ray} sources
identified with two ONIR counterparts in this catalog.
Col. (14): Expected number of false matches. See \S\ref{results} for details.
Col. (15): False-match probability. See \S\ref{results} for details.
Col. (16): Counterpart recovery rate, defined as the expected number of
true counterparts ($N_{\rm ID}-N_{\rm False}$) divided by the number of 
X-ray sources ($N_{\rm X}$).
Col. (17): The fraction of ONIR objects that are detected in the X-ray,
defined as the expected number of
true counterparts ($N_{\rm ID}-N_{\rm False}$) divided by the number of
ONIR sources in the CDF-S region ($N_{\rm o}$).
Col. (18): Number of primary counterparts selected from this catalog. See
\S\ref{method} for details.
}
\label{matchsum}
\end{deluxetable}
\end{turnpage}

\begin{deluxetable}{lcccccccccccc}

\tabletypesize{\small}
\tablewidth{0pt}

\tablecaption{List of the Primary Counterparts}

\tablehead{
\colhead{ID} &
\colhead{$\sigma_{\rm X}$} &
\colhead{ONIR ID} &
\colhead{ONIR RA} &
\colhead{ONIR Dec} &
\colhead{$\sigma_{\rm o}$} &
\colhead{Dis} &
\colhead{Dis/$\sigma$} &
\colhead{$R_{\rm c}$} &
\colhead{CAT} &
\colhead{MAG} &
\colhead{Multi} &
\colhead{Note} \\
\colhead{(1)}         &
\colhead{(2)}         &
\colhead{(3)}         &
\colhead{(4)}         &
\colhead{(5)}         &
\colhead{(6)}         &
\colhead{(7)}         &
\colhead{(8)}         &
\colhead{(9)}         &
\colhead{(10)}        &
\colhead{(11)}        &
\colhead{(12)}        &
\colhead{(13)}        
}

\startdata
1 &0.77&24484 &52.8924& $-$27.8347  &   0.1 &   0.89  &1.14& 0.88    &  GEMS   & 24.14 & 0 &     $\cdots$   \\
2&0.94&21150& 52.8990 & $-$27.8597  &    0.1 &   0.29  &0.31& 0.97  &     GEMS      & 22.27 & 0 & $\cdots$  \\  
3&0.65& 24081 &52.9171 & $-$27.7962 &   0.1&     0.44 & 0.68& 0.98 &      GEMS     & 22.26 & 0 & $\cdots$  \\   
4&0.55&0 &  0 &   0 &       0 &   0&    0 &    0& $\cdots$ & 0  & 0 & Edge   \\
\enddata
\tablecomments{
Table~2 is presented in its entirety in the electronic edition. An
abbreviated version of the table is shown here for guidance as
to its form and content. 
The full table contains 13 columns of
information on primary counterparts to the 462 \hbox{X-ray} sources.}
\label{cat}
\end{deluxetable}

\begin{deluxetable}{lcccccccccc}

\tabletypesize{\small}
\tablewidth{0pt}

\tablecaption{List of the Secondary Counterparts}

\tablehead{
\colhead{ID} &
\colhead{$\sigma_{\rm X}$} &
\colhead{ONIR ID} &
\colhead{ONIR RA} &
\colhead{ONIR Dec} &
\colhead{$\sigma_{\rm o}$} &
\colhead{Dis} &
\colhead{Dis/$\sigma$} &
\colhead{$R_{\rm c}$} &
\colhead{CAT} &
\colhead{MAG} \\
\colhead{(1)}         &
\colhead{(2)}         &
\colhead{(3)}         &
\colhead{(4)}         &
\colhead{(5)}         &
\colhead{(6)}         &
\colhead{(7)}         &
\colhead{(8)}         &
\colhead{(9)}         &
\colhead{(10)}         &
\colhead{(11)}         
}

\startdata
21&0.51 &24151 &52.9602& $-$27.8768  &   0.3 &   1.38   &2.34&     0.95    &  SIMPLE & 21.40   \\
25&0.39&35263& 52.9648 & $-$27.7648  &    0.3 &   1.56  & 3.16&  0.65  &   SIMPLE      & 21.99 \\ 
31&0.62& 23319 &52.9752 & $-$27.8349 &   0.1&     0.73 & 1.16&   0.63 &      GEMS   &21.23 \\ 
41&0.43& 30803 &  52.9906 &    $-$27.7022 &  0.1 &   0.34& 0.78& 0.98 & GEMS & 23.39 \\
\enddata
\tablecomments{
Table~3 is presented in its entirety in the electronic edition. An
abbreviated version of the table is shown here for guidance as
to its form and content.
The full table contains 11 columns of
information on the secondary counterparts of
72 \hbox{X-ray} sources (77 lines). A high 
$R_{\rm c}$ value for the secondary counterpart indicates that 
the primary and secondary counterparts were identified in different catalogs.}
\label{cat2nd}
\end{deluxetable}

\begin{deluxetable}{lccccccccc}
\tabletypesize{\small}
\tablewidth{0pt}
\tablecaption{Photometric Catalog}

\tablehead{
\colhead{ID} &
\colhead{{\it GALEX}} &
\colhead{} &
\colhead{} &
\colhead{} &
\colhead{} &
\colhead{} &
\colhead{VIMOS} &
\colhead{} &
\colhead{} \\
\colhead{} &
\colhead{NUV} &
\colhead{Err} &
\colhead{Flag} &
\colhead{FUV} &
\colhead{Err} &
\colhead{Flag} &
\colhead{$U$} &
\colhead{Err} &
\colhead{Flag} \\
\colhead{(1)}         &
\colhead{(2)}         &
\colhead{(3)}         &
\colhead{(4)}         &
\colhead{(5)}         &
\colhead{(6)}         &
\colhead{(7)}         &
\colhead{(8)}         &
\colhead{(9)}         &
\colhead{(10)}
}

\startdata
Median&23.39 &1.32&99& 23.87 & 1.66 &62& 25.81 & 3.79    &440   \\
1 &$-$26.89 &99.00&1& $-$27.25  & 99.00 &1& $-$28.53 & 99.00    &1   \\
2&23.36& 0.08 &1& $-$27.24  &  99.00&1& 99.00 &  99.00  &0 \\
3& $-$26.90 &99.00 &1& $-$27.23 & 99.00&1& 22.32 & 0.01 & 0 \\
4& 99.00 & 99.00 & 0&99.00 &99.00 &0& 99.00&  99.00 &0 \\
\enddata
\tablecomments{
Table~4 is presented in its entirety in the electronic edition. An
abbreviated version of the table is shown here for guidance as
to its form and content.
The full table contains 181 columns of photometric data 
for the 462 \hbox{X-ray} sources (Rows~2--463,
including 16 sources with no counterparts).
The first row includes the median magnitude information for each band.
All the
photometric data are given in units of AB magnitudes, and are set to ``99''
if not available; upper limits are indicated by the ``$-$'' sign.}
\label{phocat}
\end{deluxetable}

\begin{deluxetable}{lccccccccccc}
\tabletypesize{\small}
\tablewidth{0pt}
\tablecaption{Photometric Redshift Catalog}
\tablehead{
\colhead{ID} &
\colhead{Spec-z} &
\colhead{Photo-z} &
\colhead{zLower} &
\colhead{zUpper} &
\colhead{\#Detect} &
\colhead{\#Filter} &
\colhead{$\chi^2_{\rm dof}$} &
\colhead{Photo-z$_{\rm alt}$} &
\colhead{$\chi^2_{\rm dof,alt}$} &
\colhead{Op. Det.} &
\colhead{zRef} \\
\colhead{(1)}         &
\colhead{(2)}         &
\colhead{(3)}         &
\colhead{(4)}         &
\colhead{(5)}         &
\colhead{(6)}         &
\colhead{(7)}         &
\colhead{(8)}         &
\colhead{(9)}         &
\colhead{(10)}         &
\colhead{(11)}         &
\colhead{(12)}  
}

\startdata
1 & $-1.000$ & 1.102 & 1.045  & 1.114 & 13 & 16    &1.8& $-1.0$&$-1.0$&1& $-$1   \\
2 & $-1.000$ & 1.626 & 1.620  & 1.633 & 26 & 27    &23.6&$-1.0$&$-1.0$&1& $-$1   \\
3 & 2.709 & 2.714 & 2.712  & 2.715 & 26 & 28    &8.3& $-1.0$&$-1.0$&1&8   \\
4 & $-1.000$ & $-1.000$ & $-1.000$  & $-1.000$ & 0 & 0    &$-1.0$& $-1.0$&$-1.0$&0&$-$1   \\
\enddata
\tablecomments{
Table~5 is presented in its entirety in the electronic edition. An
abbreviated version of the table is shown here for guidance as
to its form and content.
The full table contains 12 columns of information
for the 462 \hbox{X-ray} sources (including 16 sources with no counterparts
and six stars).}
\label{zcat}
\end{deluxetable}

\end{document}